\documentclass[preprint,showpacs,preprintnumbers,amsmath,amssymb]{revtex4}

\usepackage{graphicx}
\usepackage{epsfig}
\usepackage{bm}
\usepackage{amsfonts}
\usepackage{subfigure}
\begin{document}

\title{Structure formation and generalized second law
of thermodynamics in some viable $f(R)$-gravity models}

\author{S. Asadzadeh} \email{ss.asadzadeh@gmail.com}

\author{M.S. Khaledian} \email{MS.Khaledian@gmail.com}

\author{K. Karami} \email{KKarami@uok.ac.ir}

\affiliation{Department of Physics, University of Kurdistan,
Pasdaran St., Sanandaj, Iran}

\date{\today}

\begin{abstract}
Here, we investigate the growth of matter density perturbations as
well as the generalized second law (GSL) of thermodynamics in the
framework of $f(R)$-gravity. We consider a spatially flat FRW
universe filled with the pressureless matter and radiation which is
enclosed by the dynamical apparent horizon with the Hawking
temperature. For some viable $f(R)$ models containing the
Starobinsky, Hu-Sawicki, Exponential, Tsujikawa and AB
models, we first explore numerically the evolution of some
cosmological parameters like the Hubble parameter, the Ricci scalar,
the deceleration parameter, the density parameters and the equation
of state parameters. Then, we examine the validity of GSL and obtain
the growth factor of structure formation. We find that for the
aforementioned models, the GSL is satisfied from the early times to
the present epoch. But in the farther future, the GSL for
the all models is violated. Our numerical results also show that
for the all models, the growth factor for larger structures like the
$\Lambda$CDM model fit the data very well.
\end{abstract}



\maketitle

\clearpage

\section{Introduction}
The observed accelerated expansion of the universe, as evidenced by
a host of cosmological data such as supernovae Ia (SNeIa)
\cite{SNeIa}, cosmic microwave background (CMB) \cite{WMAP,Planck},
large scale structure (LSS) \cite{LSS}, etc., came as a great
surprise to cosmologists. The present accelerated phase of the
universe expansion reveals new physics missing from our universe's
picture, and it constitutes the fundamental key to understand the
fate of the universe.

There are two representative approaches to
explain the current acceleration of the universe. One is to
introduce ``dark energy'' (DE) \cite{Padmanabhan} in the framework
of general relativity (GR). The other is to consider a theory of
modified gravity (MG), such as $f(R)$ gravity, in which the
Einstein-Hilbert action in GR is generalized from the Ricci scalar
$R$ to an arbitrary function of the Ricci scalar \cite{Sotiriou1}.
Here, we will focus on the later approach.

In \cite{Sotiriou2}, it
was shown that a $f(R)$ model with negative and positive powers of
Ricci curvature scalar $R$ can naturally combine the
inflation at early times and the cosmic acceleration at late times.
It is actually possible for viable $f(R)$ models for late
time acceleration to include inflation by adding $R^2$ term.
Therefore, it is natural to consider combined $f(R)$ models which
describe both primordial and present DE using one $f(R)$ function,
albeit one containing two greatly different characteristic energy
scales \cite{AB,Motohashi}. In \cite{Sobouti}, it was pointed out
that the $f(R)$-gravity can also serve as dark matter (DM). In
\cite{Khaledian}, a set of $f(R)$-gravity models corresponding to
different DE models were reconstructed. Although a great variety of
$f(R)$ models have been proposed in the literature, most of them is
not perfect enough. An interesting
feature of the $f(R)$ theories is the fact that the gravitational
constant in $f(R)$-gravity, varies with length scale as well as with
time \cite{Raccanelli}-\cite{deltam}. Thus the evolution of the matter density perturbation,
$\delta_{\rm m}\equiv\delta\rho_{\rm m}/\rho_{\rm m}$, in this
theory is affected by the effective Newton coupling constant,
$G_{\rm eff}$, and it is scale dependent too. Therefore, the matter
density perturbation is a crucial tool to distinguish MG from DE
model in GR, in particular the standard $\Lambda${CDM} model.

On the other hand, the connection between gravity and thermodynamics
is one of surprising features of gravity which was first reinforced
by Jacobson \cite{Jacobson}, who associated the Einstein field
equations with the Clausius relation in the context of black hole
thermodynamics. This idea was also extended to the cosmological
context and it was shown that the Friedmann equations in the
Einstein gravity \cite{Cai05} can be written in the form of the
first law of thermodynamics (the Clausius relation). The equivalence
between the first law of thermodynamics and the Friedmann equation
was also found for $f(R)$-gravity \cite{Akbar12}. Besides the first
law, the generalized second law (GSL) of gravitational
thermodynamics, which states that entropy of the fluid inside the
horizon plus the geometric entropy do not decrease with time, was
also investigated in $f(R)$-gravity \cite{kka1}. The GSL of
thermodynamics in the accelerating universe driven by DE or MG has
been also studied extensively in the literature
\cite{Izquierdo1}-\cite{Geng}.

All mentioned in above motivate us to investigate the growth of
matter density perturbations in a class of metric $f(R)$ models and
see scale dependence of growth factor. Additionally, we are
interested in examining the validity of GSL in some viable
$f(R)$-gravity models. The structure of this paper is as follows. In
Sec. \ref{fR}, within the framework of $f(R)$-gravity we consider a
spatially flat Friedmann-Robertson-Walker (FRW) universe filled with
the pressureless matter and radiation. In Sec. \ref{LSS}, we study
the growth rate of matter density perturbations in $f(R)$-gravity.
In Sec. \ref{GSL}, the GSL of thermodynamics on the dynamical
apparent horizon with the Hawking temperature is explained. In Sec.
\ref{cosmoevo}, the cosmological evolution of $f(R)$ models is
illustrated. In Sec. \ref{viablefR}, the viability conditions for
$f(R)$ models are discussed. In addition, some viable $f(R)$ models
containing the Starobinsky, Hu-Sawicki, Exponential, Tsujikawa and
AB models are introduced. In Sec. \ref{NR}, we give
numerical results obtained for the evolution of some cosmological
parameters, the GSL and the growth of structure formation in the
aforementioned $f(R)$ models. Section \ref{con} is devoted to
conclusions.

\section{$f(R)$-gravity framework}\label{fR}

Within the framework of $f(R)$-gravity, the modified
Einstein-Hilbert action in the Jordan frame is given by
\cite{Sotiriou1}
\begin{equation}\label{f action}
S_{\rm J} = \int {\sqrt { - g} }~{\rm d}^4
x\left[\frac{{f(R)}}{{16\pi G }} + L_{\rm matter}\right],
\end{equation}
where $G$, $g$, $R$ and $L_{\rm matter}$ are the gravitational
constant, the determinant of the metric $g_{\mu\nu}$, the Ricci
scalar and the lagrangian density of the matter inside the universe,
respectively. Also $f(R)$ is an arbitrary function of the Ricci
scalar.

Varying the action (\ref{f action}) with respect to $g_{\mu\nu}$
yields
\begin{equation}\label{Equ F1}
F G_{\mu\nu} = 8\pi G T^{(\rm m)}_{\mu\nu}-\frac{1}{2}g_{\mu\nu}(R
F-f)+\nabla_\mu \nabla_\nu F-g_{\mu\nu} \square F.
\end{equation}
Here $F={\rm d}f/{\rm d}R$,
$G_{\mu\nu}=R_{\mu\nu}-\frac{1}{2}Rg_{\mu\nu}$ and $T^{(\rm
m)}_{\mu\nu}$ is the energy-momentum tensor of the matter. The
gravitational field equations (\ref{Equ F1}) can be rewritten in the
standard form as \cite{action,TX}
\begin{equation}\label{Equ F2}
 G_{\mu\nu} = 8\pi G \big(T^{(\rm m)}_{\mu\nu}+T^{(\rm D)}_{\mu\nu}\big),
\end{equation}
with
\begin{equation}\label{TX}
 8\pi G T^{(\rm D)}_{\mu\nu}=(1-F) G_{\mu\nu}-\frac{1}{2}g_{\mu\nu}(R F-f)+\nabla_\mu \nabla_\nu F-g_{\mu\nu} \square F.
\end{equation}

For a spatially flat FRW metric, taking $T_\nu^{\mu(m)}={\rm
diag}(-\rho,p,p,p)$ in the prefect fluid form, then the set of field
equations (\ref{Equ F2}) reduce to the modified Friedmann equations
in the framework of $f(R)$-gravity as \cite{Capozziello2}
\begin{eqnarray}
3H^2 &=& 8\pi G(\rho+\rho_{\rm D}),\label{FiEq1}\\
2\dot{H}&=& -8\pi G(\rho+\rho_{\rm D}+p+p_{\rm D}),\label{FiEq2}
\end{eqnarray}
where
\begin{eqnarray}
8\pi G\rho_{\rm D} &=&\frac{1}{2}\big(RF-f\big)-3H\dot{F}+3H^2\big(1-F\big),\label{Xdensity}\\
8\pi Gp_{\rm D}&=&
\left[\frac{-1}{2}\big(RF-f\big)+\ddot{F}+2H\dot{F}-(1-F)\big(2\dot{H}+3H^2\big)\right],\label{Xpressure}
\end{eqnarray}
with
\begin{equation}\label{R}
R =6(\dot{H}+2H^2).
\end{equation}
Here $H=\dot{a}/a$ is the Hubble parameter. Also $\rho_{\rm D}$ and
$p_{\rm D}$ are the curvature contribution to the energy density and
pressure which can play the role of DE. Also $\rho=\rho_{\rm
BM}+\rho_{\rm DM}+\rho_{\rm rad}$ and $p=p_{\rm rad}=\rho_{\rm
rad}/3$ are the energy density and pressure of the matter inside the
universe, consist of the pressureless baryonic and dark matters as
well as the radiation. On the whole of the paper, the dot and the
subscript $R$ denote the derivatives with respect to the cosmic time
$t$ and the Ricci scalar $R$, respectively.

The energy conservation laws are still given by
\begin{eqnarray}
&&\dot{\rho}_{\rm m}+3H\rho_{\rm m}=0,\label{con1}\\
&&\dot{\rho}_{\rm rad}+4H\rho_{\rm rad}=0,\label{con2}\\
&&\dot{\rho}_{\rm D}+3H(\rho_{\rm D}+p_{\rm D})=0,\label{con3}
\end{eqnarray}
where $\rho_{\rm m}=\rho_{\rm BM}+\rho_{\rm DM}$. From Eqs.
(\ref{con1}) and (\ref{con2}) one can find
\begin{equation}\label{ro m)}
\rho=\frac{\rho_{\rm m_0}}{a^3}+\frac{\rho_{\rm rad_0}}{a^4},
\end{equation}
where $\rho_{\rm m_0}=\rho_{\rm BM_0}+\rho_{\rm DM_0}$ and
$\rho_{\rm rad_0}$ are the present values of the energy densities of
matter and radiation. We also choose $a_0=1$ for the recent value of
the scale factor.

Using the usual definitions of the density parameters
\begin{eqnarray}\label{OmegaDif}
 \Omega_{\rm m} = \frac{\rho_{\rm m}}{\rho_{\rm c}}=\frac{8\pi G\rho_{\rm m_0}}{3H^2a^3},~~~
 \Omega_{\rm rad} =  \frac{\rho_{\rm rad}}{\rho_{\rm c}}= \frac{8\pi G\rho_{\rm rad_0}}{3H^2a^4},~~~
 \Omega_{\rm D} =  \frac{\rho_{\rm D}}{\rho_{\rm c}}=\frac{8\pi G\rho_{\rm D}}{3H^2},
\end{eqnarray}
in which $\rho_{\rm c}=3H^2/(8\pi G)$ is the critical energy
density, the modified Friedmann equation (\ref{FiEq1}) takes the
form
\begin{equation}\label{omega1}
1=\Omega_{\rm m}+\Omega_{\rm rad}+\Omega_{\rm D}.
\end{equation}
From the energy conservation (\ref{con3}), the equation of state
(EoS) parameter due to the curvature contribution is defined as
\begin{equation}\label{w D}
\omega_{\rm D}=\frac{p_{\rm D}}{\rho_{\rm
D}}=-1-\frac{\dot{\rho}_{\rm D}}{3H\rho_{\rm D}}.
\end{equation}
Using the modified Friedmann equations (\ref{FiEq1}) and
(\ref{FiEq2}), the effective EoS parameter is obtained as
\begin{equation}\label{w eff}
\omega_{\rm eff}=\frac{p+p_{\rm D}}{\rho+\rho_{\rm
D}}=-1-\frac{2\dot{H}}{3H^2}.
\end{equation}
Also the two important observational cosmographic parameters called
the deceleration $q$ and the jerk $j$ parameters, respectively
related to $\ddot{a}$ and $\dddot{a}$, are given by
\cite{cosmography}
\begin{eqnarray}
  q &=& -\frac{\ddot{a}}{aH^2}=-1-\frac{\dot{H}}{H^2}=1-\frac{R}{6H^2}, \label{q} \\
  j &=& \frac{\dddot{a}}{aH^3}=1-\frac{\dot{H}}{H^2}+\frac{\dot{R}}{6H^3}=2+q+\frac{\dot{R}}{6H^3}.\label{j}
\end{eqnarray}

\section{Growth rate of matter density perturbations}\label{LSS}

Here, we study the evolution of the matter density contrast $\delta_{\rm m}=\delta
\rho_{\rm m}/\rho_{\rm m}$ in
$f(R)$-gravity. To this aim, we consider the linear
scalar perturbations around a flat FRW background in the Newtonian
(longitudinal) gauge as
\begin{equation}\label{metric}
    {\rm d}s^{2}=-(1+2\Psi){\rm d}t^{2}+a^2(t)(1+2\Phi){\rm d}x^2,
\end{equation}
with two scalar potentials $\Psi$ and $\Phi$ describing the
perturbations in the metric. In this gauge, the matter density
perturbation $\delta_{\rm m}$ and the perturbation of $\delta F(R)$
obey the following equations in the Fourier space \cite{Hwang,
Tsujik}
\begin{equation}\label{delta-m}
    \ddot{\delta}_{\rm m}+\left(2H+\frac{\dot{F}}{2F}\right)\dot{\delta}_{\rm
    m}
    -\frac{8\pi G \rho_{\rm m}}{2F}\delta_{\rm m}=\frac{1}{2F}\left[\left(-6H^2+\frac{k^2}{a^2}\right)\delta
    F+3H\dot{\delta F}+3\ddot{\delta F}\right],
\end{equation}
\begin{equation}\label{delta-F}
    \ddot{\delta F}+3H\dot{\delta F}
    +\left(\frac{k^2}{a^2}+\frac{F}{3F_{\rm R}}-\frac{R}{3}\right)\delta
    F=\frac{8\pi G}{3}\rho_{\rm m}\delta_{\rm m}+\dot{F}\dot{\delta}_{\rm m},
\end{equation}
where $k$ is the comoving wave number. For the modes deep inside the
Hubble radius (i.e. $k^2/a^2\gg H^2$), we have $|\dot{F}|\ll HF$ and $\ddot{\delta F}\ll H
\dot{\delta F} \ll H^2$, hence the evolution of matter density
 contrast $\delta_{\rm m}$ reads \cite{Starobinsky,Tsu}
\begin{equation}\label{delta}
    \ddot{\delta}_{\rm m}+2H\dot{\delta}_{\rm m}-4\pi
    G_{\rm eff}\rho_{\rm m}\delta_{\rm m}=0,
\end{equation}
where
\begin{equation}\label{G_eff}
G_{\rm
eff}=\frac{G}{F}\left[\frac{4}{3}-\frac{1}{3}\frac{M^{2}a^{2}}{k^{2}+M^{2}a^{2}}\right],
\end{equation}
and $M^{2}=\frac{F}{3F_{\rm R}}$. The fraction of effective
gravitational constant to the Newtonian one, i.e. $G_{\rm eff}/G$,
is defined as screened mass function in the literature
\cite{Rahvar}. Equation (\ref{G_eff}) obviously shows that the
screened mass function is the time and scale dependent parameter.

With the help of new variable namely $g(a)=\delta_{\rm m}/a$ which
parameterizes the growth of structure in the matter, Eq.
(\ref{delta}) becomes
\begin{equation}
\frac{{\rm d}^2g}{{\rm
d}\ln{a^2}}+\left(4+\frac{\dot{H}}{{H}^{2}}\right)\frac{{\rm d}
g}{{\rm d}\ln{a}}
+\left(3+\frac{\dot{H}}{{H}^{2}}-\frac{4{\pi}G_{\rm eff}\rho_{\rm
m}}{H^2}\right)g=0.\label{eqg1}
\end{equation}
In general, there is no analytical solution to this
equation. But in \cite{Yokoyama} for an asymptotic form of viable
$f(R)$ models at high curvature regime given by $f(R)=R+R^{-n}$
where $n>-1$, an analytic solution for density perturbations in the
matter component during the matter dominated stage was obtained in
terms of hypergeometric functions. In what follows we solve the
differential equation (\ref{eqg1}), numerically. To this aim, the
natural choice for the initial conditions are $g(a_{\rm m})=1$ and
$\frac{\rm d g}{\rm d\ln a}\mid_{a=a_{\rm m}}=0$, where $a_{\rm
m}=1/(1+z_{\rm m})$ should be taken during the matter era, because
for the matter dominated universe, i.e. $H^2=8\pi G\rho_{\rm m}/3$
and $G_{\rm eff}/G=1$, the solution of Eq. (\ref{delta}) yields
$\delta_{\rm m}=a$.
  The growth factor is defined as \cite{Peebles}
\begin{equation}
f(z)=\frac{{\rm d}\ln\delta_{\rm m}}{{\rm d}\ln
a}=-(1+z)\,\frac{{\rm d}\ln\delta_{\rm m}}{{\rm d}z},
\end{equation}
which is an observational parameter. In the
present work, we obtain the evolution of linear perturbations
relevant to the matter spectrum for the scales; $k=0.1, 0.01, 0.001$
$h~{\rm Mpc}^{-1}$, where $h$ corresponds to the Hubble parameter
today.

\section{Generalized second law of thermodynamics}\label{GSL}

Here, we are interested in examining the validity of the GSL of
gravitational thermodynamics for a given $f(R)$ model. According to
the GSL, entropy of the matter inside the horizon beside the entropy
associated with the surface of horizon should not decrease during
the time \cite{Cai05}. As demonstrated by Bekenstein, this
law is satisfied by black holes in contact with their radiation
\cite{Bekenstein}. The entropy of the matter containing the
pressureless matter and radiation inside the horizon is given by the
Gibbs' equation \cite{Izquierdo1}
\begin{equation}\label{gibbs}
T_{\rm A}{\rm d}S={\rm d}E+p {\rm d}V.
\end{equation}
Taking time derivative of Eq.
(\ref{gibbs}) and using the energy equations
(\ref{con1})-(\ref{con2}) as well as the Friedmann equations
(\ref{FiEq1})-(\ref{FiEq2}) one can find
\begin{equation}\label{matterantropy}
T_{\rm A}\dot{S}=\frac{\tilde{r}^2_{\rm A}}{2G}\left(
\dot{\tilde{r}}_{\rm A}-H\tilde{r}_{\rm A}\right)\left(-
2\dot{H}+H\frac{{\rm d}}{{\rm d}t}-\frac{ {\rm d}^2}{{\rm
d}t^2}\right)F,
\end{equation}
where $\tilde{r}_{\rm A}=(H^2+\frac{K}{a^2})^{-1/2}$ and $T_{\rm
A}=\frac{1}{2\pi \tilde{r}_{\rm A}}\big(1-\frac{\dot{\tilde{r}}_{\rm
A}}{2H\tilde{r}_{\rm A}}\big)$ are the dynamical apparent horizon
 and Hawking temperature, respectively. The horizon entropy in $f(R)$-gravity is given by $S_{\rm
A}=\frac{AF}{4G}$ \cite{Wald}, where $A=4\pi\tilde{r}^2_{\rm A}$ is
the area of the apparent horizon. Taking the time derivative of
$S_{\rm A}$ one can get the evolution of horizon entropy as
\begin{equation}\label{EM}
T_{\rm A}\dot{S}_{\rm A} =\frac{1}{4GH} \left( 2H\tilde{r}_{\rm
A}-\dot{\tilde{r}}_{\rm A} \right) \left(
\frac{2\dot{\tilde{r}}_{\rm A}}{\tilde{r}_{\rm A}}+\frac{{\rm
d}}{{\rm d}t} \right)F.
\end{equation}
Now we can calculate the GSL due to different contributions of the
matter and horizon. Adding Eqs. (\ref{matterantropy}) and
(\ref{EM}), one can get the GSL in $f(R)$-gravity as \cite{kka1}
 \begin{equation}\label{TASdotflat}
T_{\rm A}\dot{S}_{\rm
tot}=\frac{1}{4GH^{4}}\left[2\dot{H}^{2}F-\dot{H}H\dot{F}+2(\dot{H}+H^{2})\ddot{F}\right],
\end{equation}
where $S_{\rm tot}=S+S_{\rm A}$. Note that Eq. (\ref{TASdotflat})
shows that the validity of the GSL, i.e. $T_{\rm A}\dot{S}_{\rm
tot}\geq 0$, depends on the $f(R)$-gravity model. For the
Einstein gravity ($F=1$), one can immediately find that the GSL
(\ref{TASdotflat}) reduces to
\begin{equation}\label{GSL-R}
T_{\rm A}\dot{S}_{\rm tot}=\frac{\dot{H}^{2}}{2GH^{4}}\geq 0,
\end{equation}
which shows that the GSL is always fulfilled throughout history of
the universe.

\section{Cosmological evolution}\label{cosmoevo}

Here, we recast the differential equations governing the evolution
of the universe in dimensionless form which is more suitable for
numerical integration. To do so, following \cite{Luisa} we use the
dimensionless quantities
\begin{equation}\label{dimless1}
  \bar{t} = H_0t,~~~\bar{H} = \frac{H}{H_0},~~~\bar{R} =
  \frac{R}{H^{2}_0},
  \end{equation}

\begin{equation}\label{dimless2}
   \bar{f} = \frac{f}{H^{2}_0},~~~\bar{F}=F,~~~\bar{F}_{\rm R} = \frac{F_{\rm R}}{H^{-2}_0},~~~
   \bar{F}_{\rm RR }= \frac{F_{\rm RR}}{H^{-4}_0},
 \end{equation}
where $H_0$ is the Hubble parameter today. With the help of the
above definitions and using
\begin{equation}
  \frac{\rm d}{{\rm d}\bar{t}} = -\bar{H}(1+z)\frac{\rm d}{{\rm d}z},
\end{equation}
one can rewrite the modified Friedmann equation (\ref{FiEq1}) as
follows
\begin{equation}\label{H2}
\bar{H}^2= \Omega_{\rm m_0}\big[(1+z)^3+\chi (1+z)^4
\big]+(\bar{F}-1)\big[\bar{H}^2-(1+z)\bar{H}
\bar{H}'\big]-\frac{1}{6}\big(\bar{f}-\bar{R}\big)+(1+z)\bar{H}^2
\bar{F}_R
   \bar{R}',
\end{equation}
where $\chi=\rho_{\rm rad_0}/\rho_{\rm m_0}=\Omega_{\rm
rad_0}/\Omega_{\rm m_0}$ and prime `$'$' denotes a derivative with
respect to the cosmological redshift $z=\frac{1}{a}-1$.

To solve Eq. (\ref{H2}) we introduce new variables as \cite{HS}:
\begin{equation}\label{yH}
    y_{\rm H}:=\frac{\rho_{\rm D}}{\rho_{\rm m_0}}=\frac{\bar{H}^2}{\Omega_{\rm m_0}}-(1+z)^3-\chi(1+z)^4,
\end{equation}
and
\begin{equation}\label{yR}
    y_{\rm R}:=\frac{\bar{R}}{\Omega_{\rm m_0}}-3(1+z)^3.
\end{equation}
Taking the derivative of both sides of Eqs. (\ref{yH}) and
(\ref{yR}) with respect to redshift $z$ yield
\begin{equation}\label{yHprime}
    -(1+z){y'_{\rm H}}=\frac{1}{3}y_{\rm R}-4y_{\rm H},
   \end{equation}
\begin{eqnarray}\label{yRprime}
      -(1+z){y'_{\rm R}}&=&9(1+z)^3-\frac{1}{\bar{H}^2\bar{F}_{\rm R}}\left\{y_{\rm H}+\frac{1}{6\Omega_{\rm
    m_0}}(\bar{f}-\bar{R})\right.\nonumber\\
  &&\left.-(\bar{F}-1)\left[\frac{y_{\rm R}}{6}-y_{\rm H}-\frac{1}{2}\Big((1+z)^3+2\chi(1+z)^4\Big)\right]\right\}.
\end{eqnarray}
Finally, inserting Eq. (\ref{yRprime}) into the derivative of Eq.
(\ref{yHprime}) gives a second differential equation governing
$y_{\rm H}(z)$ as \cite{Bamba}
\begin{equation}\label{yH dif}
    (1+z)^2{y''_{\rm H}}+J_1(1+z){y'_{\rm H}}+J_2y_{\rm H}+J_3=0,
\end{equation}
where
\begin{equation}\label{Js1}
  J_1 = -3-\left(\frac{1-\bar{F}}{6\bar{H}^2\bar{F}_{\rm R}}\right), \\
 \end{equation}
\begin{equation}\label{Js2}
    J_2 = \frac{2-\bar{F}}{3\bar{H}^2\bar{F}_{\rm R}}, \\
  \end{equation}
\begin{equation}\label{Js3}
    J_3 = -3(1+z)^3-\frac{1}{6\bar{H}^2\bar{F}_{\rm R}}\left[(1-\bar{F})\Big((1+z)^3+2\chi(1+z)^4\Big)+\frac{1}{3\Omega_{\rm m_0}}(\bar{R}-\bar{f})\right].
\end{equation}
Equation (\ref{yH dif}) cannot be solved analytically. Hence, we
need to solve it numerically. To do so, we use the two initial
conditions $y_{\rm H}(z_{\rm i})=3$ and $y'_{\rm H}(z_{\rm i})=0$
which come from the $\Lambda$CDM approximation of $f(R)$ model in
high curvature regime. Notice $z_{\rm i}$ is the proper redshift in
which we have $RF_{\rm R}(z_{\rm i})\leq10^{-13}$.

With the help of Eqs. (\ref{OmegaDif}), (\ref{w D}), (\ref{w eff})
and (\ref{yH}) one can obtain the evolutionary behaviors of the
matter density parameter, $\Omega_{\rm m}(z)$, DE density parameter,
$\Omega_{\rm D}(z)$, EoS parameter of DE, $\omega_{\rm D}(z)$, and
effective EoS parameter, $\omega_{\rm eff}(z)$, in terms of $y_{\rm
H}$ and its derivatives as follows
\begin{equation}\label{Om m}
    \Omega_{\rm m}(z)=\frac{(1+z)^3}{y_{\rm H}+(1+z)^3+\chi(1+z)^4},
\end{equation}
\begin{equation}\label{Om D}
    \Omega_{\rm D}(z)=\frac{y_{\rm H}}{y_{\rm H}+(1+z)^3+\chi(1+z)^4},
\end{equation}
\begin{equation}\label{qu4}
  \omega_{\rm D}(z) = -1+\frac{1+z}{3}\left(\frac{y'_{\rm H}}{y_{\rm H}}\right),
\end{equation}
\begin{equation}\label{qu3}
   \omega_{\rm eff}(z) = -1+\frac{(1+z)}{3}\left[\frac{ y'_{\rm H}+3(1+z)^2+4\chi(1+z)^3}{ y_{\rm H}+(1+z)^3+\chi(1+z)^4}\right].
\end{equation}
Also from Eqs. (\ref{q}), (\ref{j}) and (\ref{yH}) one can get the
evolutions of the deceleration and jerk parameters as
\begin{equation}\label{qu1}
    q(z) = -1+\frac{(1+z)}{2}\left[\frac{ y'_{\rm H}+3(1+z)^2+4\chi(1+z)^3}{ y_{\rm H}+(1+z)^3+\chi(1+z)^4}\right],
\end{equation}
\begin{equation}\label{qu2}
    j(z) = 1+\frac{(1+z)}{2}\left[\frac{(1+z)y''_{\rm H}-2 y'_{\rm H}+4\chi(1+z)^3}{ y_{\rm H}+(1+z)^3+\chi(1+z)^4}\right].
\end{equation}

\section{Viable $f(R)$-gravity models}\label{viablefR}

Since we are intersected in investigating the growth of structure
formation and examining the GSL in $f(R)$-gravity, hence in what
follows we consider some viable $f(R)$ models including the
Starobinsky, Hu-Sawicki, Exponential, Tsujikawa and AB
models.

\subsection{Starobinsky Model}

The Starobinsky $f(R)$ model is as follows \cite{Starobinsky}
\begin{equation}\label{Staro}
    f(R)=R+\lambda R_{\rm
s}\left[\left(1+\frac{R^2}{R^2_{\rm s}}\right)^{-n}-1\right],
\end{equation}
where $n>0$, $\lambda$ and $R_{\rm s}$ are constant parameters of
the model. Following \cite{pstaro}, we take $n=2$ and $\lambda=1$.
Note that in the high $z$ regime ($z\simeq z_{\rm i}$) we have
$R/R_{\rm s} \gg 1$. This yields the $f(R)$ model (\ref{Staro}) to
behave like the $\Lambda$CDM model, i.e. $f(R)=R-2\Lambda$.
Consequently, the constant parameter $R_{\rm s}$ is obtained as
$R_{\rm s}=18\Omega_{\rm m_0}H^2_0/\lambda$.

\subsection{Hu-Sawicki Model}

This model was reconstructed based on the local observational data
and presented by Hu and Sawicki \cite{HS} as
\begin{equation}\label{HS}
    f(R)=R-\frac{c_1 R_{\rm s}\big(\frac{R}{R_{\rm s}}\big)^n}{c_2\big(\frac{R}{R_{\rm s}}\big)^n+1},
\end{equation}
where $n>0$, $c_1, c_2$ and $R_{\rm s}$ are constants of the model.
For this model we take $n=4$, $c_1=1.25\times10^{-3},
c_2=6.56\times10^{-5}$ \cite{Luisa}, and obtain $R_{\rm
s}=18c_2\Omega_{\rm m_0 }H^2_0/c_1$.

\subsection{Exponential Model}

This model is defined by the following function \cite{Bamba},
\begin{equation}\label{EXP}
    f(R)=R-\beta R_{\rm s}\left(1-e^{-\frac{R}{R_{\rm s}}}\right),
\end{equation}
where $\beta$ and $R_{\rm s}$ are two constants of the model. Here
$R_{\rm s}$ corresponds to the characteristic curvature modification
scale. Here we take $\beta=1.8$ \cite{Bamba} and obtain $R_{\rm
s}=18\Omega_{\rm m_0}H^2_0/\beta$.

\subsection{Tsujikawa Model}

This model was originally presented in \cite{Tsujik} as
\begin{equation}\label{Tsu}
    f(R)=R-\lambda R_{\rm s} \tanh{\left(\frac{R}{R_{\rm s}}\right)},
\end{equation}
where $\lambda$ and $R_{\rm s}$ are the model parameters. For this
model we obtain $R_{\rm s}=18\Omega_{\rm m_0}H^2_0/\lambda$ and set
$\lambda=1$ \cite{Bamba2}.

\subsection{AB Model}
This model was proposed by Appleby and Battye \cite{AB,AB1}
as
\begin{equation}\label{AB1}
    f(R)=\frac{R}{2}+\frac{\epsilon}{2}\log\left[\frac{\cosh\big(\frac{R}{\epsilon}-b\big)}{\cosh(b)}\right],
\end{equation}
where $b$ is a dimensionless constant and $\epsilon=R_{\rm
s}/\big[b+\log(2\cosh b)\big]$. The constant $R_{\rm s}$ can be
obtained at high curvature regime when the AB $f(R)$ model
(\ref{AB1}) behaves like the $\Lambda$CDM model, i.e.
$f(R)=R-2\Lambda$. This gives $$ R_{\rm s}= \frac{-36\;\Omega_{\rm
m_0}H^2_0\big[b+\log(2\cosh b)\big]}{\log\big(\frac{1-\tanh
b}{2}\big)}.$$ Here, we also set $b=1.4$.

\section{Numerical results}\label{NR}

Here to solve Eq. (\ref{yH dif}) numerically, we choose the
cosmological parameters $\Omega_{\rm m_0}=0.24$, $\Omega_{\rm
D_0}=0.76$ and $\Omega_{\rm rad_0}=4.1\times10^{-5}$. As we have
already mentioned, we use the two suitable initial conditions
$y_{\rm H}(z_{\rm i})=3$ and $y'_{\rm H}(z_{\rm i})=0$, in which
$z_{\rm i}$ is obtained where $RF_{\rm R}\rightarrow 10^{-13}$. For
the Starobinsky, Hu-Sawicki, Exponential, Tsujikawa and AB
$f(R)$ models we obtain $z_{\rm i}=$15.61, 13.12, 3.66, 3.52 and
3.00, respectively.

In addition, to study the growth rate of matter density
perturbations, we numerically solve Eq. (\ref{eqg1}) with the
initial conditions $g(z_{\rm m})=1$ and $({\rm d}g/{\rm d}\ln
a)|_{z_{\rm m}}=0$, in which $z_{\rm m}$ is obtained where
$\Omega_{\rm m}(z_{\rm m})=1$. For the aforementioned models we
obtain $z_{\rm m}=$14, 13, 12, 14 and 14.36, respectively.

\begin{figure}[h]
\begin{minipage}[b]{1\textwidth}
\subfigure{ \includegraphics[width=.40\textwidth]%
{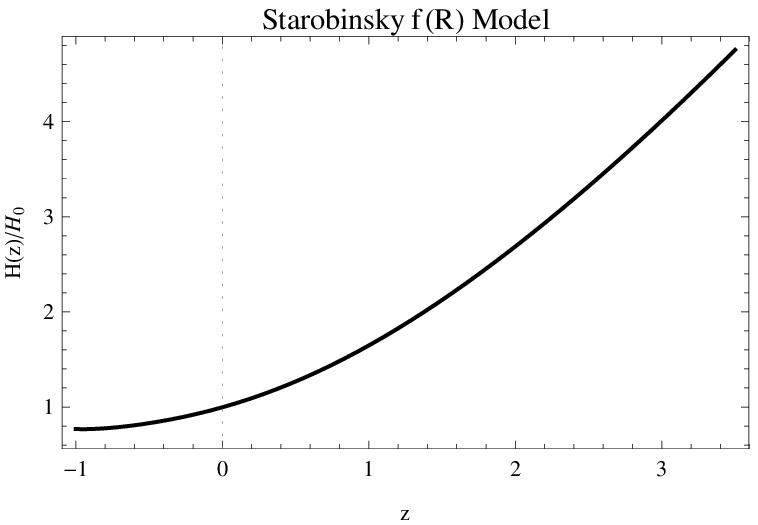}} \hspace{.2cm}
\subfigure{ \includegraphics[width=.40\textwidth]%
{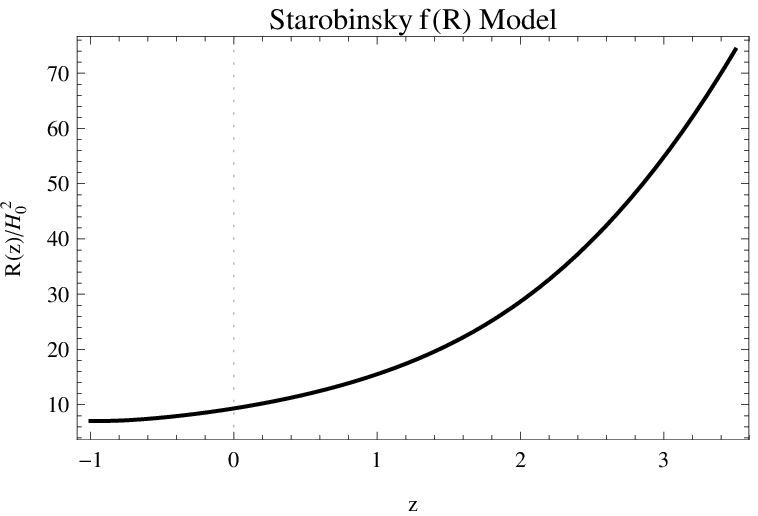}}
\end{minipage}
\begin{minipage}[b]{1\textwidth}
\subfigure{ \includegraphics[width=.40\textwidth]%
{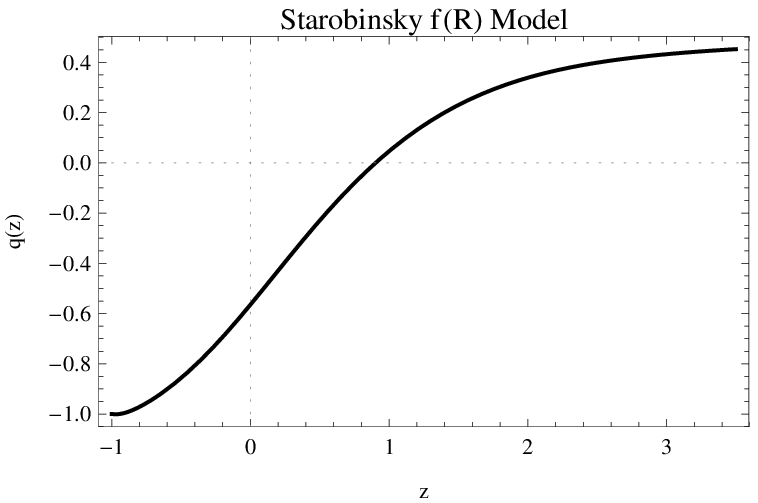}} \hspace{.2cm}
\subfigure{ \includegraphics[width=.40\textwidth]%
{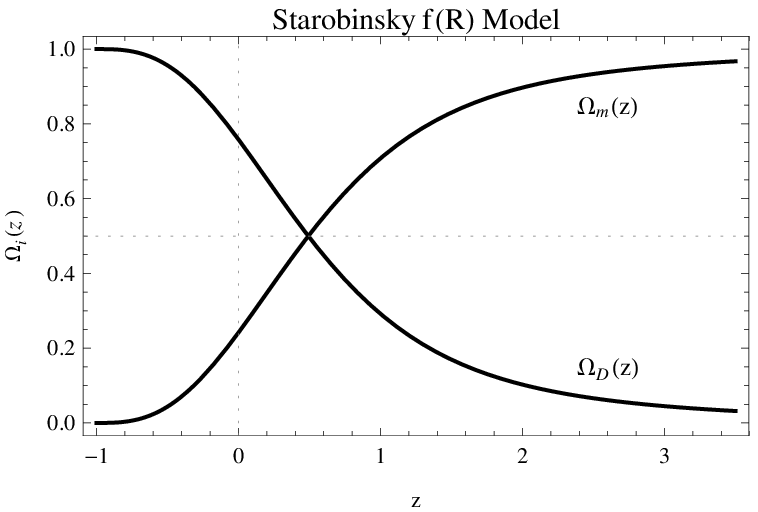}}
\end{minipage}
\begin{minipage}[b]{1\textwidth}
\subfigure{ \includegraphics[width=.40\textwidth]%
{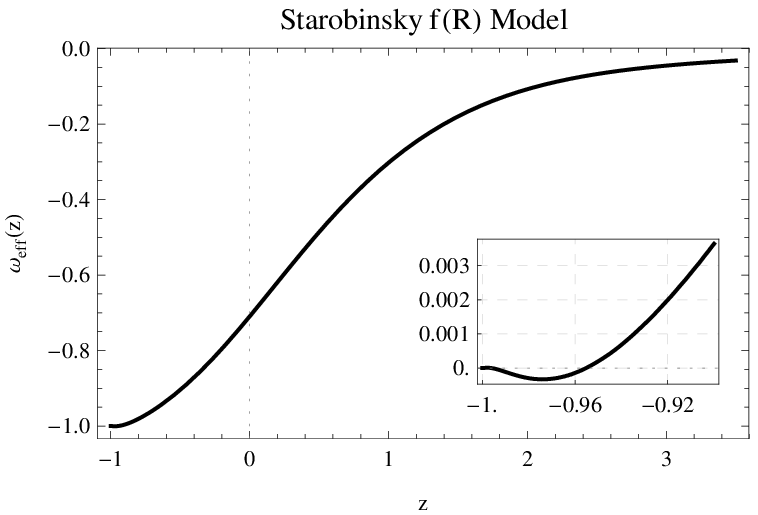}} \hspace{.2cm}
\subfigure{ \includegraphics[width=.40\textwidth]%
{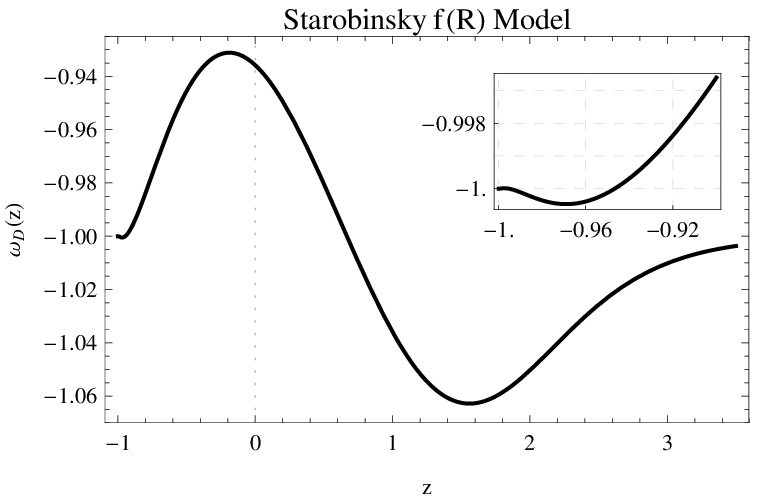}}
\end{minipage}
\begin{minipage}[b]{1\textwidth}
\subfigure{ \includegraphics[width=.40\textwidth]%
{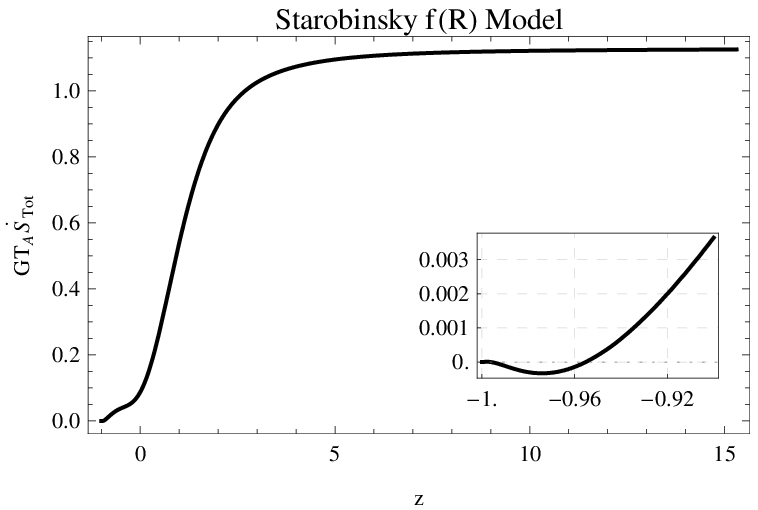}}\hspace{.05cm}
\subfigure{ \includegraphics[width=.42\textwidth]%
{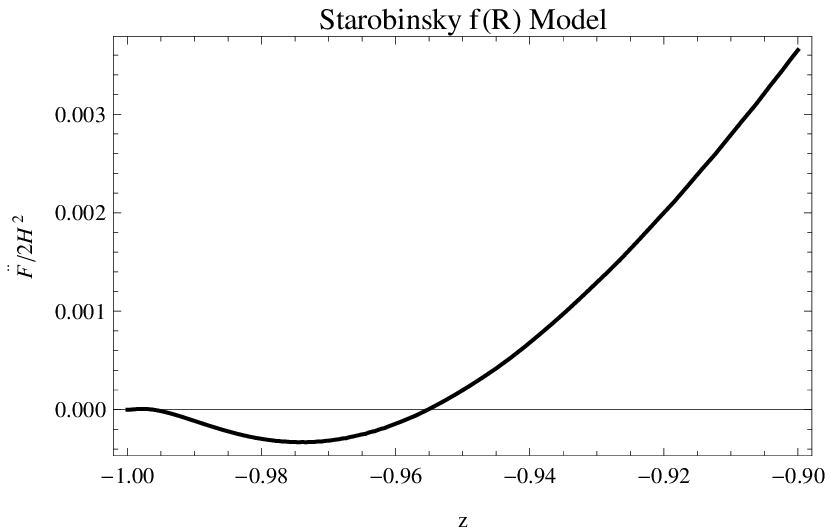}}
\end{minipage}
\caption{The variations of the Hubble parameter $H/H_0$, the Ricci
scalar $R/H_0^2$, the deceleration parameter $q$, the density
parameter $\Omega_{\rm i}$, the effective EoS parameter $\omega_{\rm
eff}$ , the EoS parameter of DE $\omega_{\rm D}$, the GSL, $G T_{\rm
A}\dot{S}_{\rm tot}$ and $\frac{\ddot{F}}{2H^2}$ versus redshift $z$
for the Starobinsky model. Auxiliary parameters are $\Omega_{\rm
m_0}=0.24$, $\Omega_{\rm D_0}=0.76$, $\Omega_{\rm
rad_0}=4.1\times10^{-5}$, $\lambda=1$ and $n=2$ .} \label{ST}
\end{figure}

\begin{figure}[h]
\begin{minipage}[b]{1\textwidth}
\subfigure{ \includegraphics[width=.40\textwidth]%
{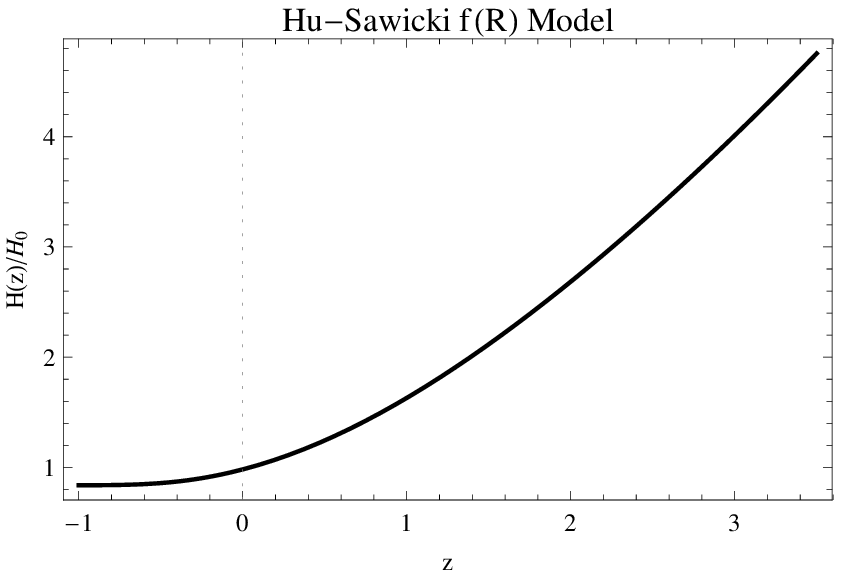}} \hspace{.2cm}
\subfigure{ \includegraphics[width=.40\textwidth]%
{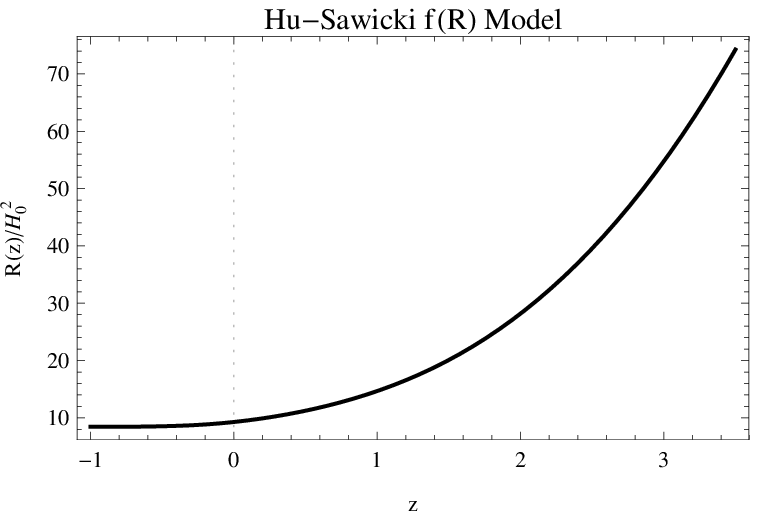}}
\end{minipage}
\begin{minipage}[b]{1\textwidth}
\subfigure{ \includegraphics[width=.40\textwidth]%
{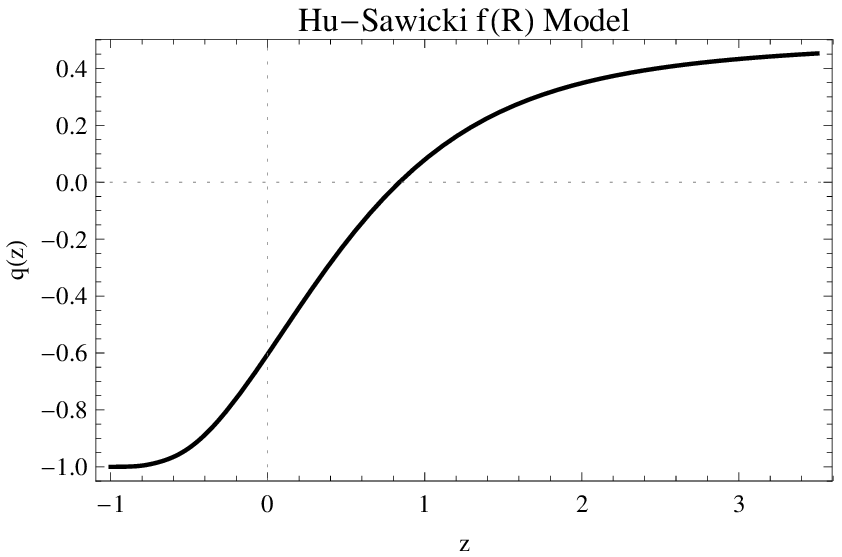}} \hspace{.2cm}
\subfigure{ \includegraphics[width=.40\textwidth]%
{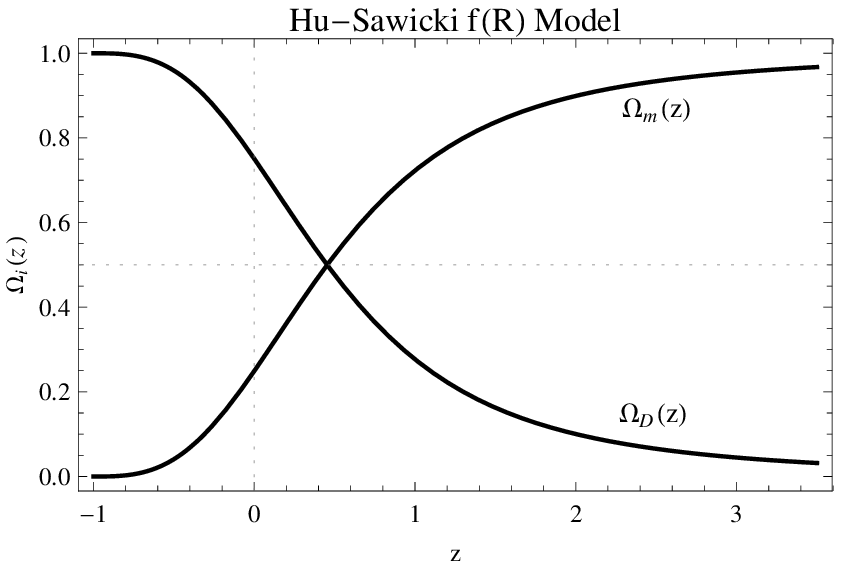}}
\end{minipage}
\begin{minipage}[b]{1\textwidth}
\subfigure{ \includegraphics[width=.40\textwidth]%
{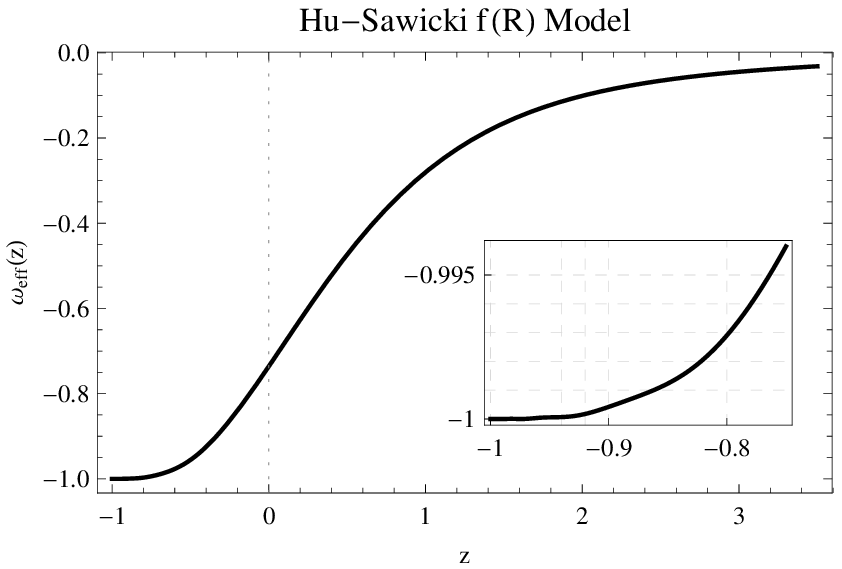}} \hspace{.2cm}
\subfigure{ \includegraphics[width=.40\textwidth]%
{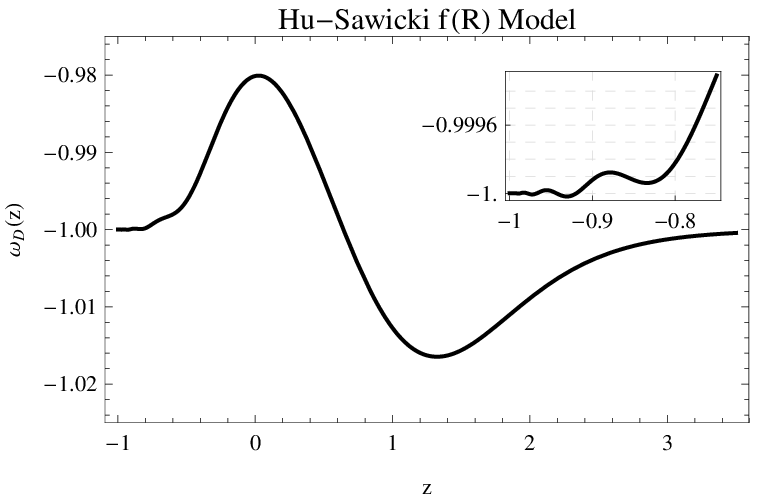}}
\end{minipage}
\begin{minipage}[b]{1\textwidth}
\subfigure{ \includegraphics[width=.40\textwidth]%
{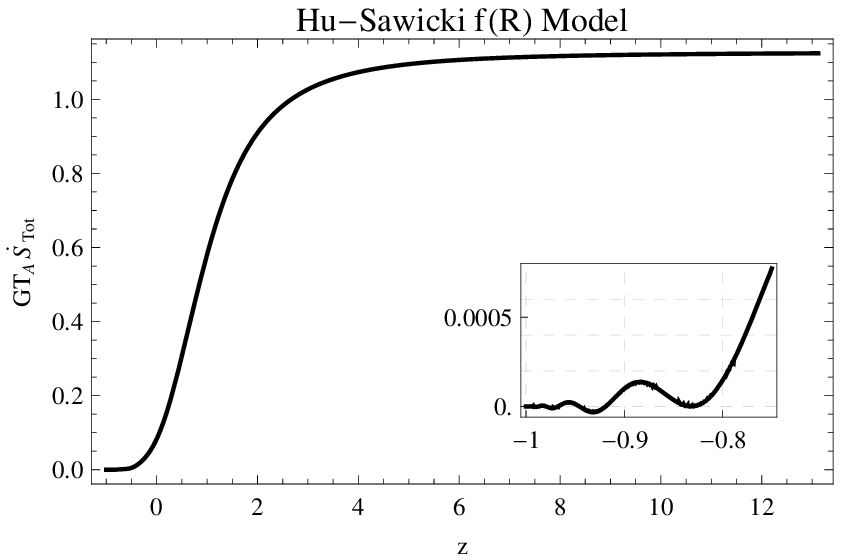}} \hspace{.02cm}
\subfigure{ \includegraphics[width=.42\textwidth]%
{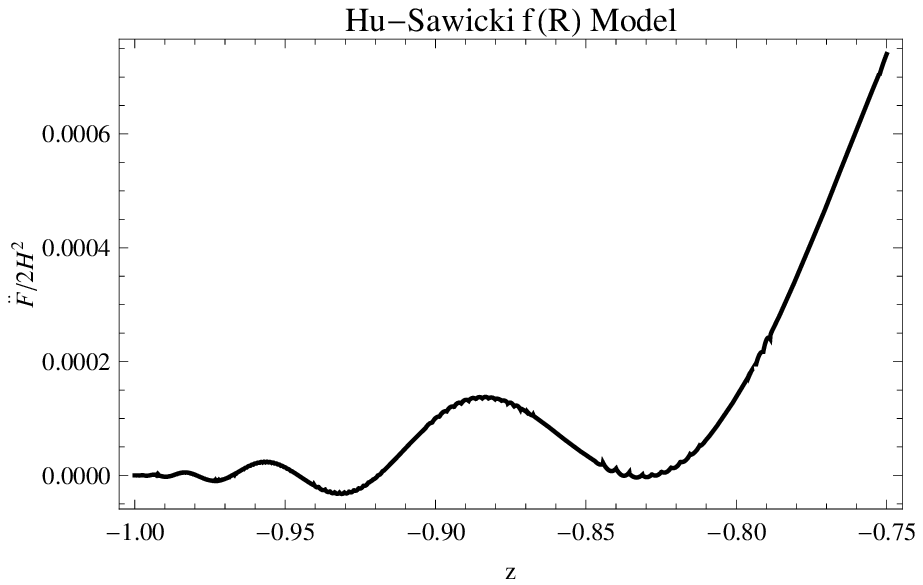}}
\end{minipage}
\caption{Same as Fig. \ref{ST} but for the Hu-Sawicki model.
Auxiliary parameters are $\Omega_{\rm m_0}=0.24$, $\Omega_{\rm
D_0}=0.76$, $\Omega_{\rm rad_0}=4.1\times10^{-5}$, $c_1=1.25\times
10^{-3}$, $c_2=6.56\times10^{-5}$ and $n=4$.} \label{Hu}
\end{figure}

\begin{figure}[h]
\begin{minipage}[b]{1\textwidth}
\subfigure{ \includegraphics[width=.40\textwidth]%
{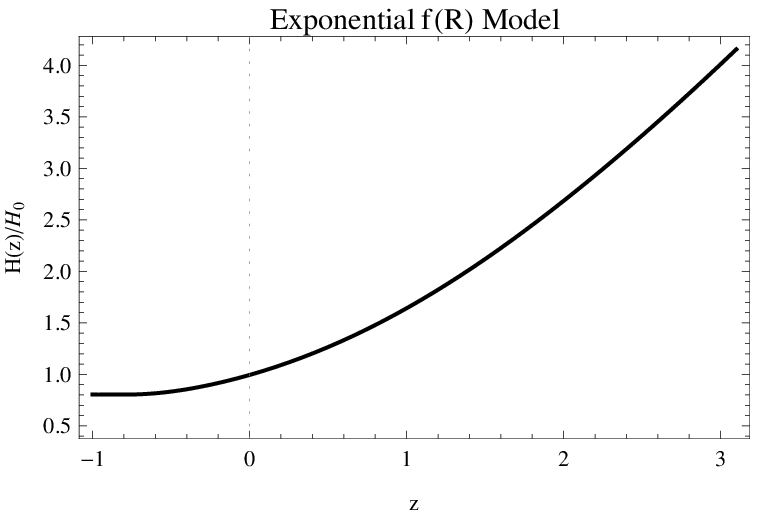}} \hspace{.2cm}
\subfigure{ \includegraphics[width=.40\textwidth]%
{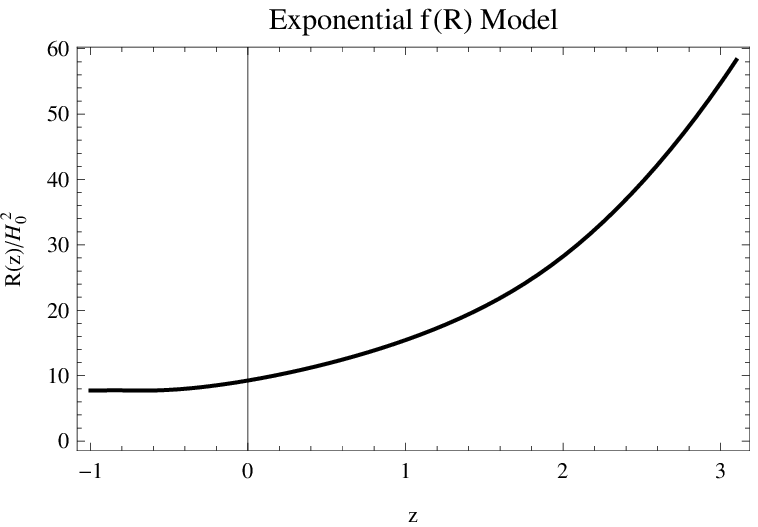}}
\end{minipage}
\begin{minipage}[b]{1\textwidth}
\subfigure{ \includegraphics[width=.40\textwidth]%
{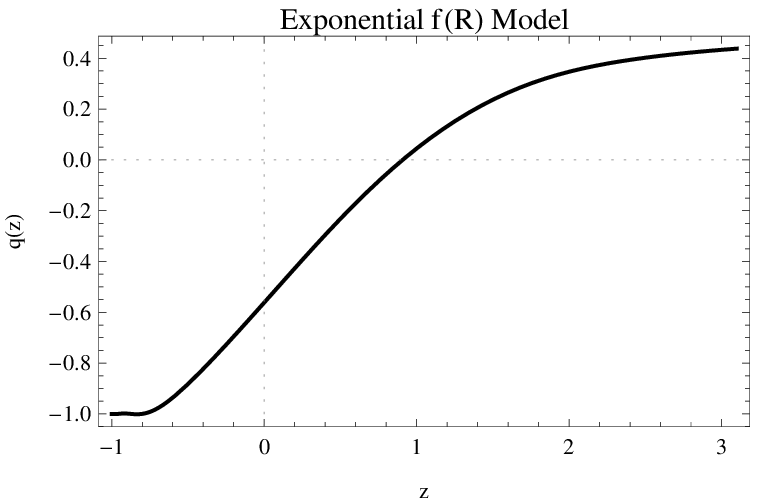}} \hspace{.2cm}
\subfigure{ \includegraphics[width=.40\textwidth]%
{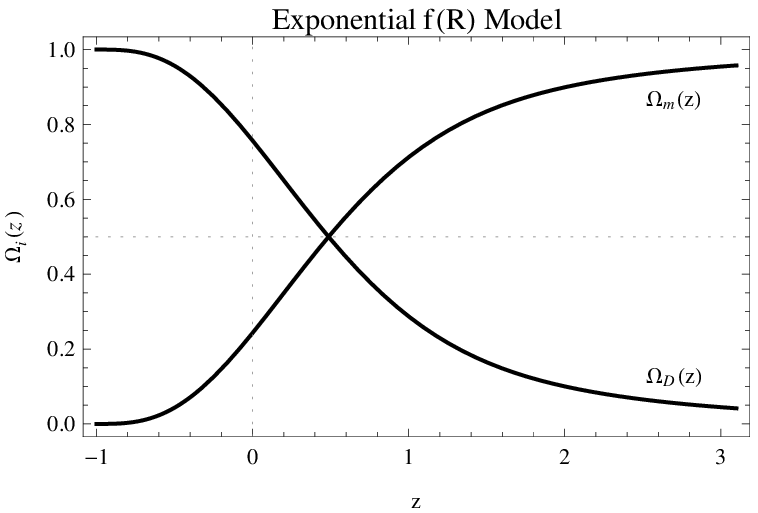}}
\end{minipage}
\begin{minipage}[b]{1\textwidth}
\subfigure{ \includegraphics[width=.40\textwidth]%
{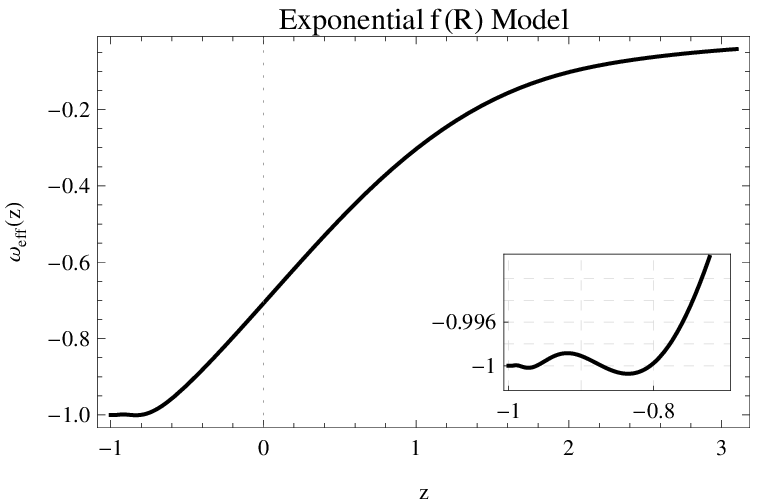}} \hspace{.2cm}
\subfigure{ \includegraphics[width=.40\textwidth]%
{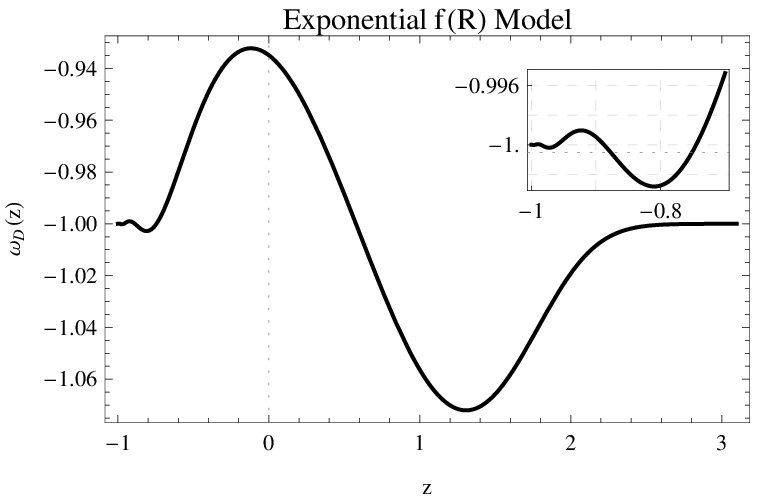}}
\end{minipage}
\begin{minipage}[b]{1\textwidth}
\subfigure{ \includegraphics[width=.40\textwidth]%
{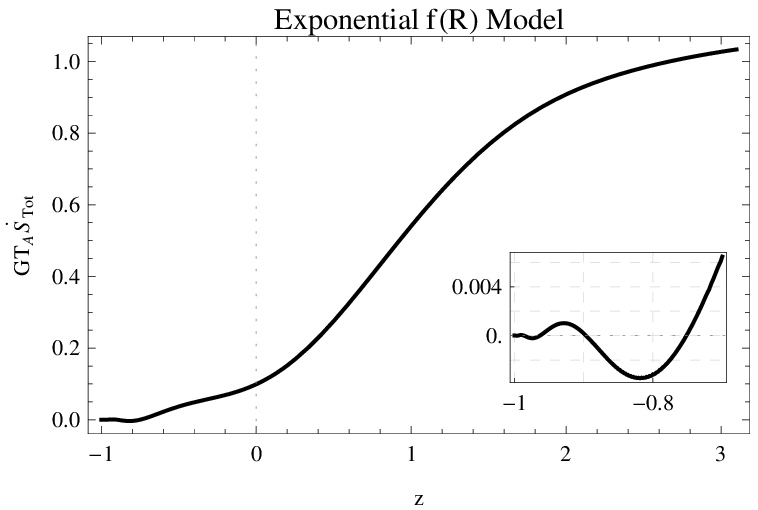}} \hspace{.05cm}
\subfigure{ \includegraphics[width=.44\textwidth]%
{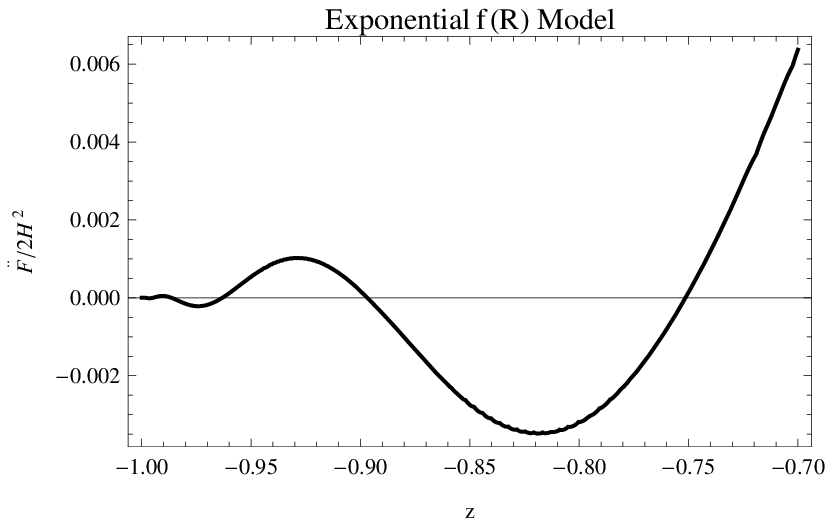}}
\end{minipage}
\caption{Same as Fig. \ref{ST} but for the Exponential model.
Auxiliary parameters are $\Omega_{\rm m_0}=0.24$, $\Omega_{\rm
D_0}=0.76$, $\Omega_{\rm rad_0}=4.1\times10^{-5}$ and $\beta=1.8$.}
\label{exp}
\end{figure}

\begin{figure}[h]
\begin{minipage}[b]{1\textwidth}
\subfigure{ \includegraphics[width=.40\textwidth]%
{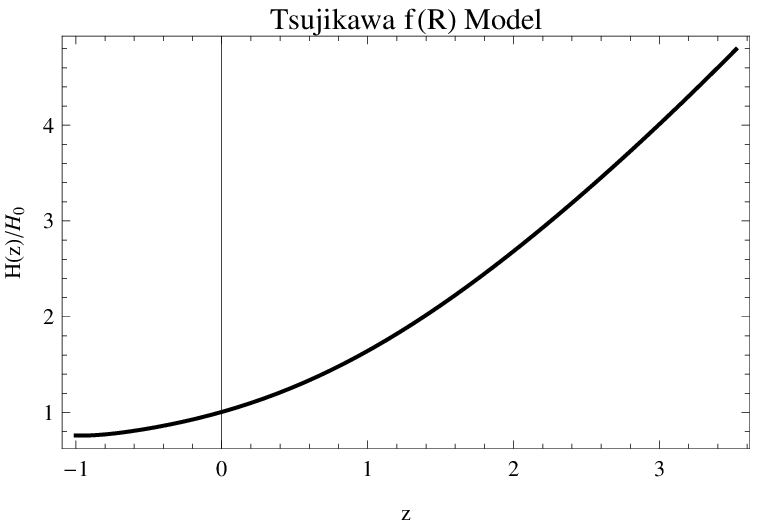}} \hspace{.2cm}
\subfigure{ \includegraphics[width=.40\textwidth]%
{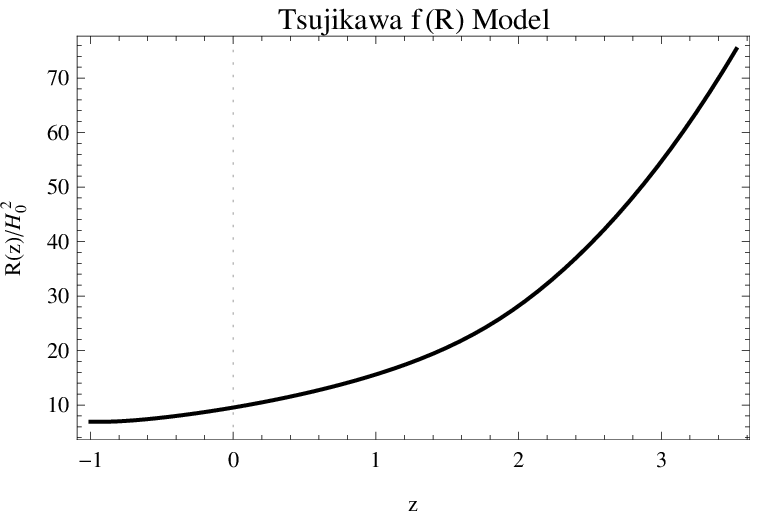}}
\end{minipage}
\begin{minipage}[b]{1\textwidth}
\subfigure{ \includegraphics[width=.40\textwidth]%
{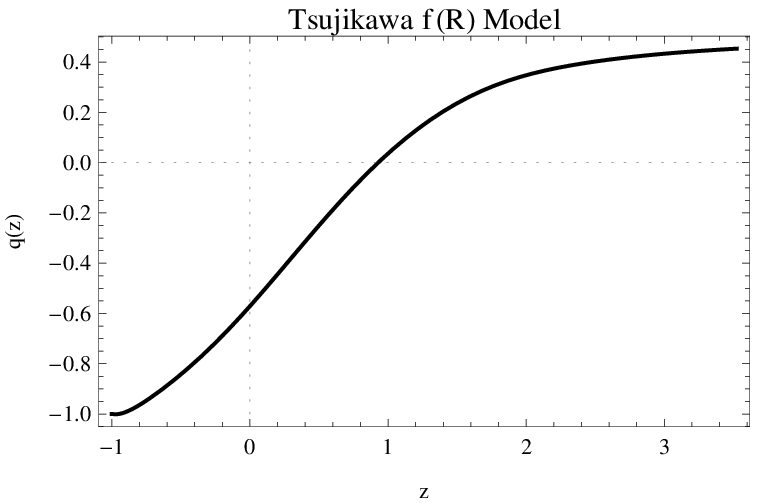}} \hspace{.2cm}
\subfigure{ \includegraphics[width=.40\textwidth]%
{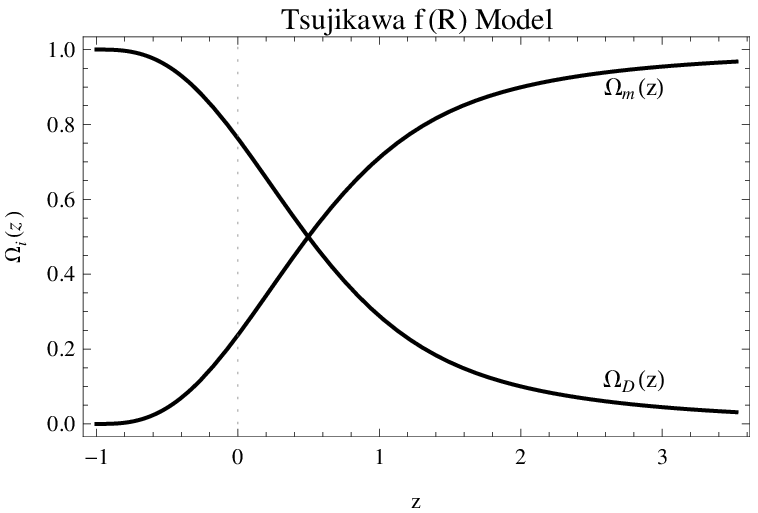}}
\end{minipage}
\begin{minipage}[b]{1\textwidth}
\subfigure{ \includegraphics[width=.40\textwidth]%
{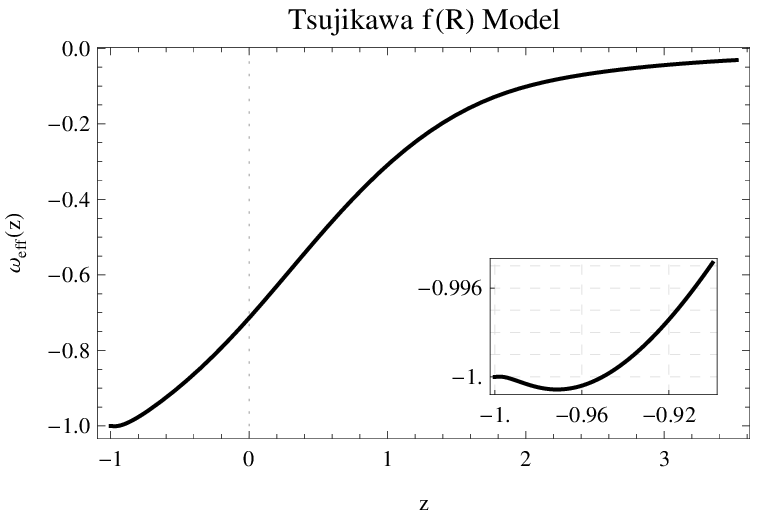}} \hspace{.2cm}
\subfigure{ \includegraphics[width=.40\textwidth]%
{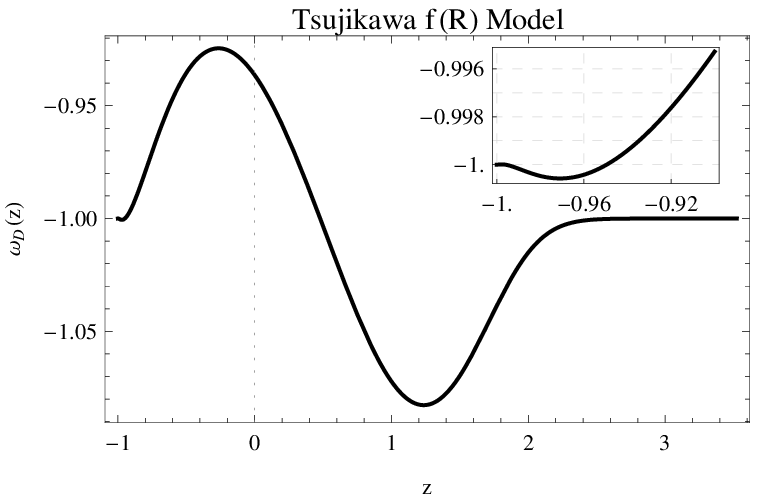}}
\end{minipage}
\begin{minipage}[b]{1\textwidth}
\subfigure{ \includegraphics[width=.40\textwidth]%
{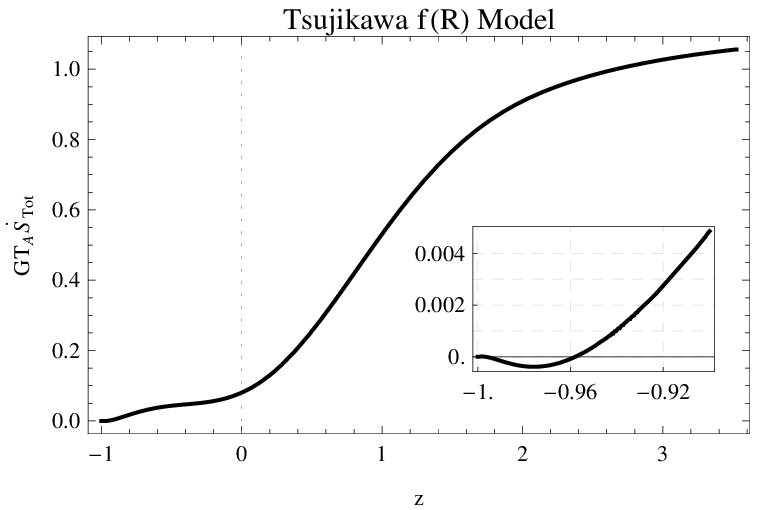}} \hspace{.05cm}
\subfigure{ \includegraphics[width=.44\textwidth]%
{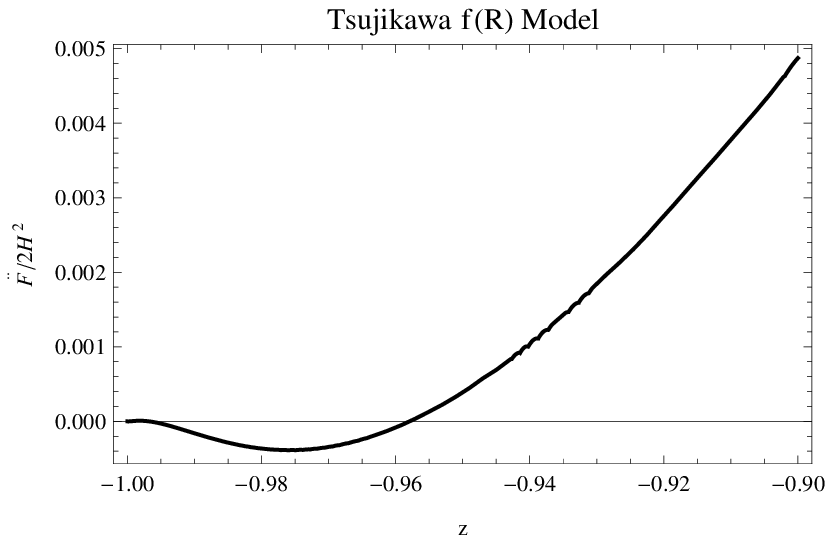}}
\end{minipage}
\caption{Same as Fig. \ref{ST} but for the Tsujikawa model.
Auxiliary parameters are $\Omega_{\rm m_0}=0.24$, $\Omega_{\rm
D_0}=0.76$, $\Omega_{\rm rad_0}=4.1\times10^{-5}$ and $\lambda=1$.}
\label{tsu}
\end{figure}

\begin{figure}[h]
\begin{minipage}[b]{1\textwidth}
\subfigure{\includegraphics[width=.4\textwidth]%
{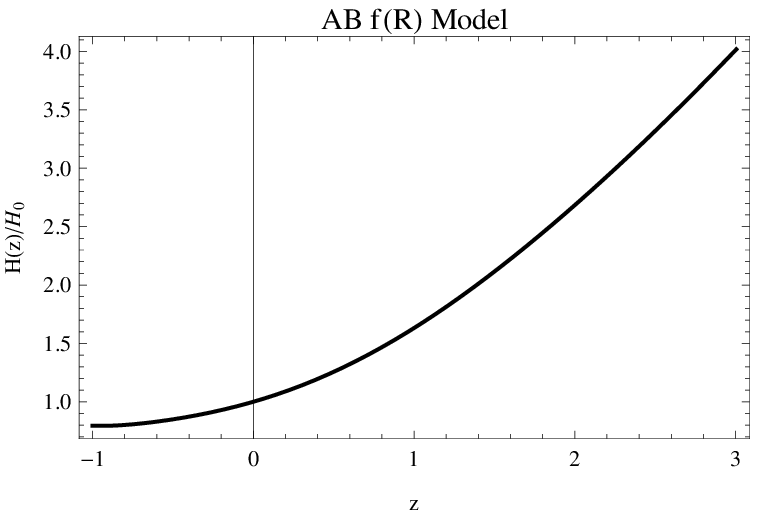}} \hspace{.05cm}
\subfigure{\includegraphics[width=.40\textwidth]%
{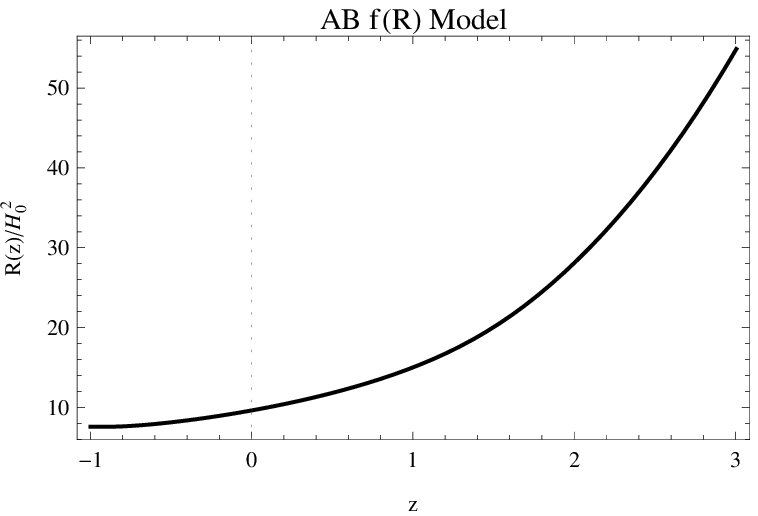}}
\end{minipage}
\begin{minipage}[b]{1\textwidth}
\subfigure{\includegraphics[width=.4\textwidth]%
{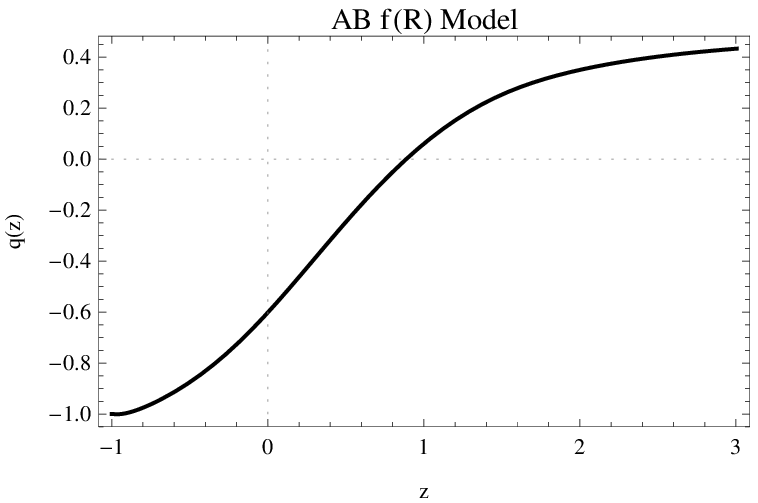}} \hspace{.05cm}
\subfigure{\includegraphics[width=.4\textwidth]%
{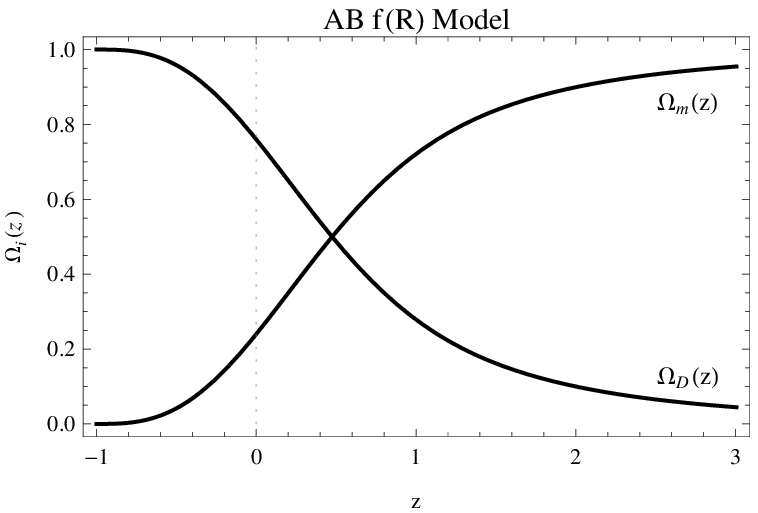}}
\end{minipage}
\begin{minipage}[b]{1\textwidth}
\subfigure{ \includegraphics[width=.4\textwidth]%
{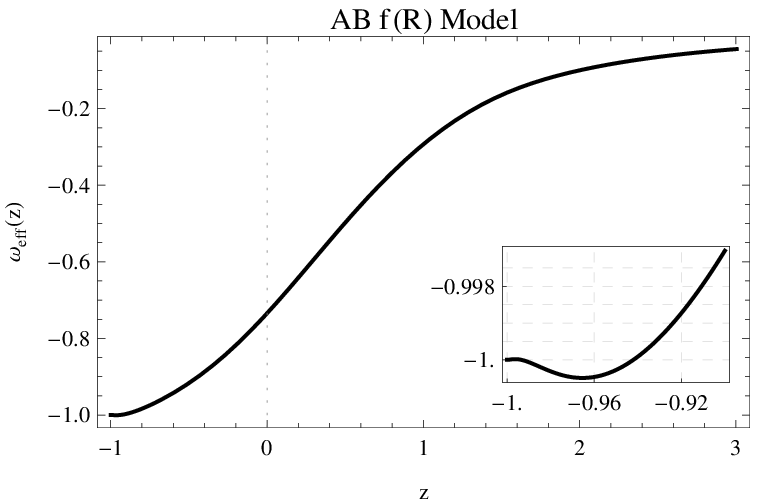}} \hspace{.05cm}
\subfigure{ \includegraphics[width=.4\textwidth]%
{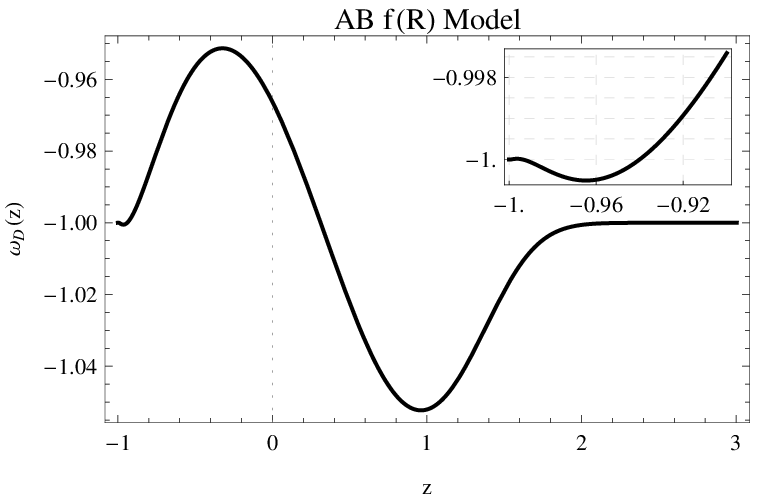}}
\end{minipage}
\begin{minipage}[b]{1\textwidth}
\subfigure{ \includegraphics[width=.4\textwidth]%
{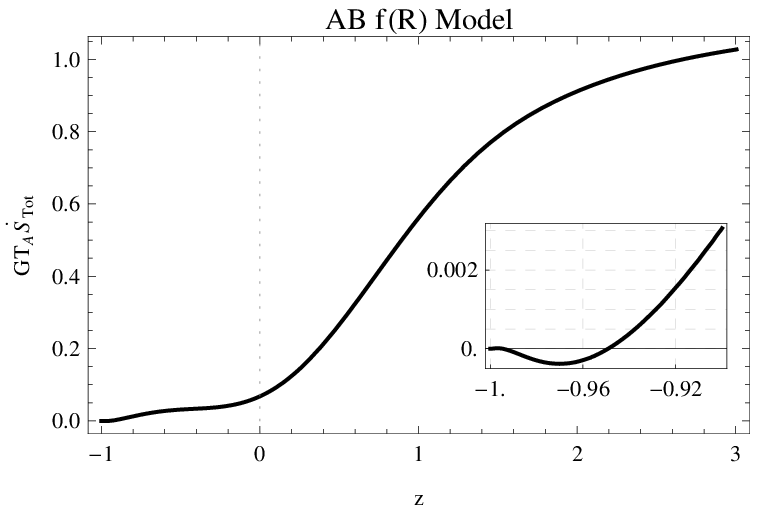}} \hspace{.05cm}
\subfigure{ \includegraphics[width=.44\textwidth]%
{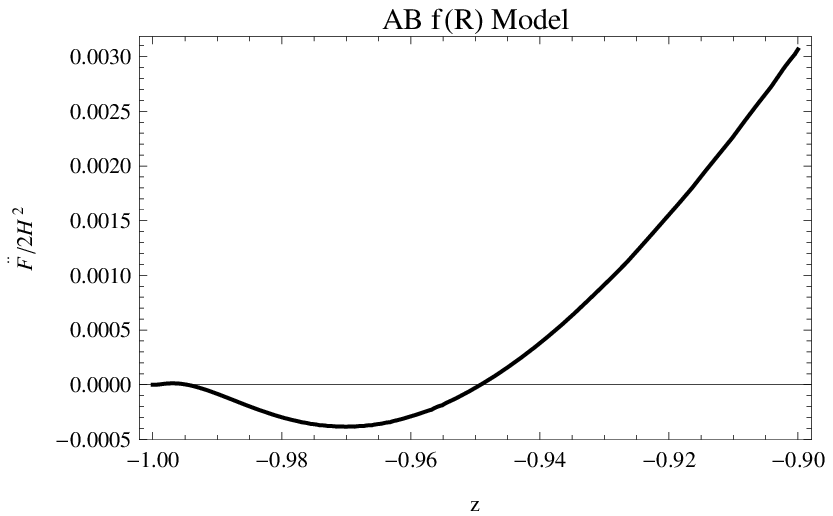}}
\end{minipage}
\caption{Same as Fig. \ref{ST} but for the AB model.
Auxiliary parameters are $\Omega_{\rm m_0}=0.24$, $\Omega_{\rm
D_0}=0.76$, $\Omega_{\rm rad_0}=4.1\times10^{-5}$ and $b=1.4$.}
\label{Ab}
\end{figure}

With the help of numerical results obtained for $y_{\rm H}(z)$ in
Eq. (\ref{yH dif}), we can obtain the evolutionary behaviors of $H$,
$R$, $q$, $\Omega_{\rm m}$, $\Omega_{\rm D}$, $\omega_{\rm eff}$,
$\omega_{\rm D}$ and GSL for our selected $f(R)$ models. The results
for the Starobinsky, Hu-Sawicki, Exponential, Tsujikawa and
AB $f(R)$ models are displayed in Figs. \ref{ST}-\ref{Ab}.
Figures show that: (i) the Hubble parameter and the Ricci scalar
decrease during history of the universe. (ii) The deceleration
parameter $q$ varies from an early matter-dominant epoch ($q=0.5$)
to the de Sitter era ($q=-1$) in the future, as expected. It also
shows a transition from a cosmic deceleration $q>0$ to the
acceleration $q<0$ in the near past. The current values of the
deceleration parameter for the Starobinsky, Hu-Sawicki, Exponential,
Tsujikawa and AB $f(R)$ models are obtained as $q_0=-0.56$,
$-0.60$, $-0.56$, $-0.57$ and $-0.60$, respectively. These are in
good agreement with the recent observational constraint
$q_0=-0.43_{-0.17}^{+0.13}~(68\%~\rm CL)$ obtained by the
cosmography \cite{CapCosmo}. (iii) The density parameters
$\Omega_{\rm D}$ and $\Omega_{\rm m}$ increases and decreases,
respectively, as $z$ decreases. (iv) The effective EoS parameter,
$\omega_{\rm eff}$, for the all models, starts from an early
matter-dominated regime (i.e. $\omega_{\rm eff}=0$) and in the late
time, $z\rightarrow -1$, it behaves like the $\Lambda$CDM model,
$\omega_{\rm eff}\rightarrow -1$. (v) The EoS parameter of
DE, $\omega_{\rm D}$, for the all models starts at the phase of a
cosmological constant, i.e. $\omega_{\rm D}=-1$, and evolves from
the phantom phase, $\omega_{\rm D}< -1$, to the non-phantom
(quintessence) phase, $\omega_{\rm D}
>-1$. The crossing of the phantom divide line $\omega_{\rm D}=-1$ occurs in the near past as well as farther future.
At late times ($z\rightarrow -1$), $\omega_{\rm D}$ approaches again
to $-1$ like the $\Lambda$CDM model. Moreover, the present values of
$\omega_{\rm D}$ for the Starobinsky, Hu-Sawicki, Exponential,
Tsujikawa and AB $f(R)$ models are obtained as $\omega_{\rm
D_0}=-0.94$, $-0.98$, $-0.93$, $-0.94$ and $-0.97$, respectively.
These values satisfy the present observational constraints
\cite{WMAP,Planck}.

(vi) The variation of the GSL shows that it holds for the
aforementioned models from early times to the present epoch. But in
the farther future, the GSL for the Starobinsky, Hu-Sawicki,
Exponential, Tsujikawa and AB $f(R)$ models is violated for
$-0.996<z<-0.955$, $-0.935<z<-0.909$, $-0.897<z<-0.751$,
$-0.997<z<-0.958$ and $-0.995<z<-0.950$, respectively. To
investigate this problem in ample detail, using Eq. (\ref{w eff}) we
rewrite Eq. (\ref{TASdotflat}) in terms of $\omega_{\rm eff}$ as
\begin{equation}\label{TASdotflat2}
T_{\rm A}\dot{S}_{\rm
tot}=\frac{1}{4G}\left[\frac{9}{2}(1+\omega_{\rm
eff})^{2}F+\frac{3}{2}(1+\omega_{\rm eff})
\frac{\dot{F}}{H}-(1+3\omega_{\rm eff})\frac{\ddot{F}}{H^2}\right],
\end{equation}
which shows that in the farther future $z\rightarrow -1$ when
$\omega_{\rm eff}\rightarrow -1$ (see Figs. \ref{ST}-\ref{Ab}) we
have
\begin{equation}\label{GSLfuture}
T_{\rm A}\dot{S}_{\rm tot}\simeq\frac{\ddot{F}}{2GH^2}.
\end{equation}
According to Eq. (\ref{GSLfuture}), the validity of GSL, i.e.
$T_{\rm A}\dot{S}_{\rm {tot}}\geq0$, depends on the sign of
$\ddot{F}$. In Figs. \ref{ST}-\ref{Ab}, we plot the variation of
$\ddot{F}/(2H^2)$ versus $z$ in the farther future for the selected
$f(R)$ models. Figures confirm that when the sign of $\ddot{F}$
changes from positive to negative due to the dominance of DE over
non-relativistic matter then the GSL is violated. Although the parameters
used for each model in Figs. \ref{ST}-\ref{Ab} are the viable ones,
by more fine tuning the model parameters the GSL can be held. For
instance, in AB $f(R)$ model by choosing the model parameter as
$b=1.3$, the GSL is always satisfied from early times to the late
cosmological history of the universe.

\begin{figure}[h]
\begin{minipage}[b]{1\textwidth}
\subfigure{ \includegraphics[width=.48\textwidth]%
{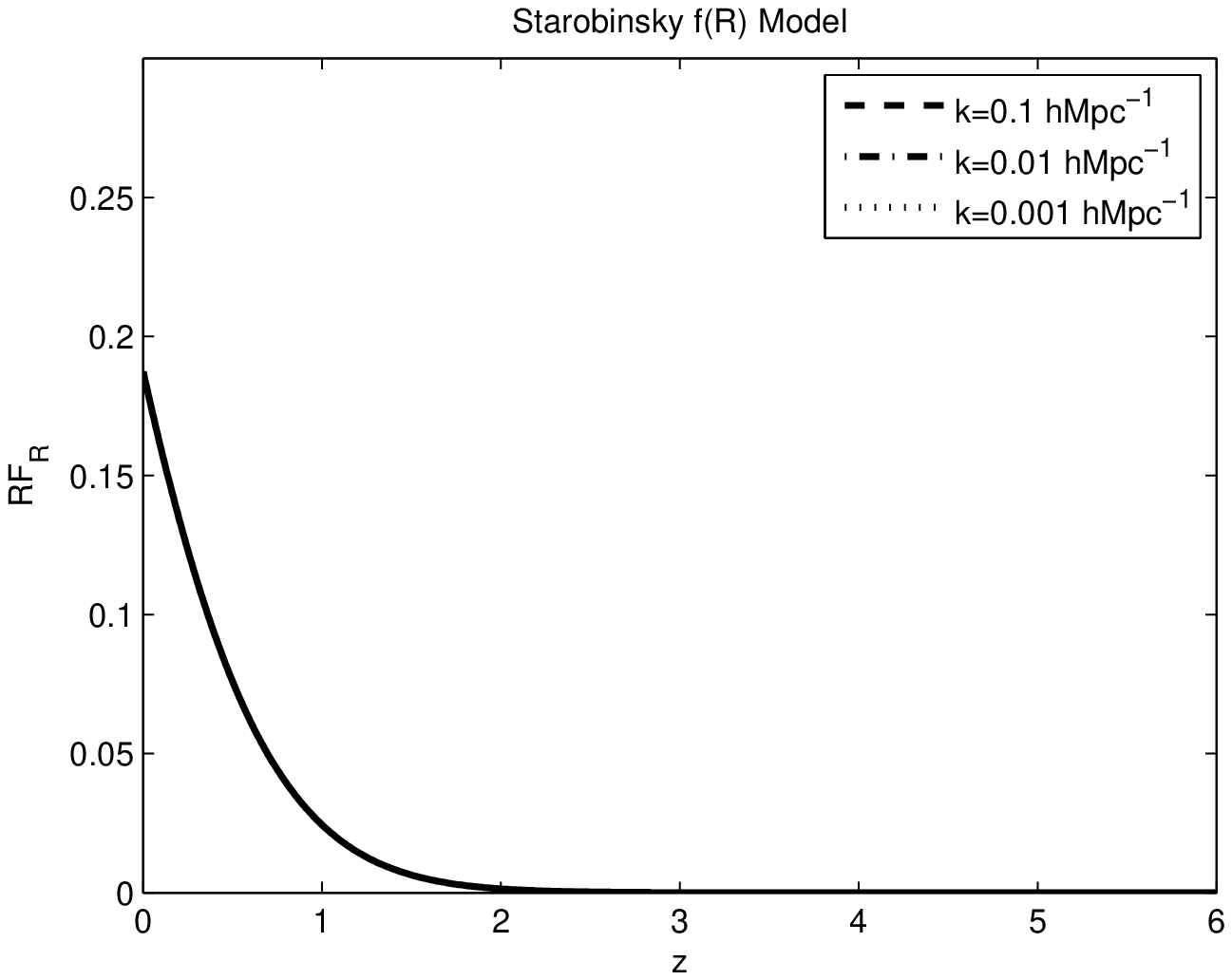}} \hspace{.1cm}
\subfigure{ \includegraphics[width=.48\textwidth]%
{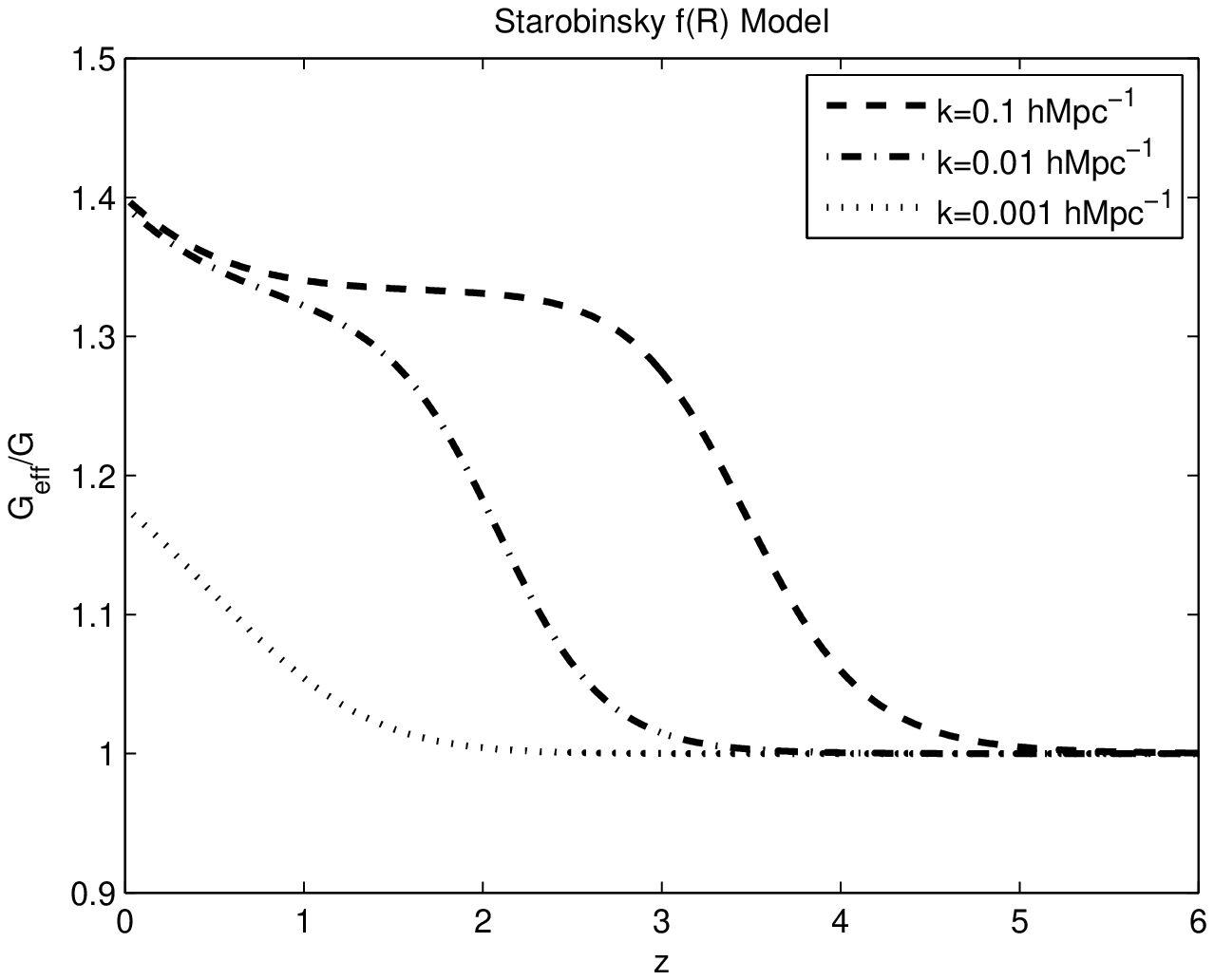}}
\end{minipage}
\begin{minipage}[b]{1\textwidth}
\subfigure{ \includegraphics[width=.48\textwidth]%
{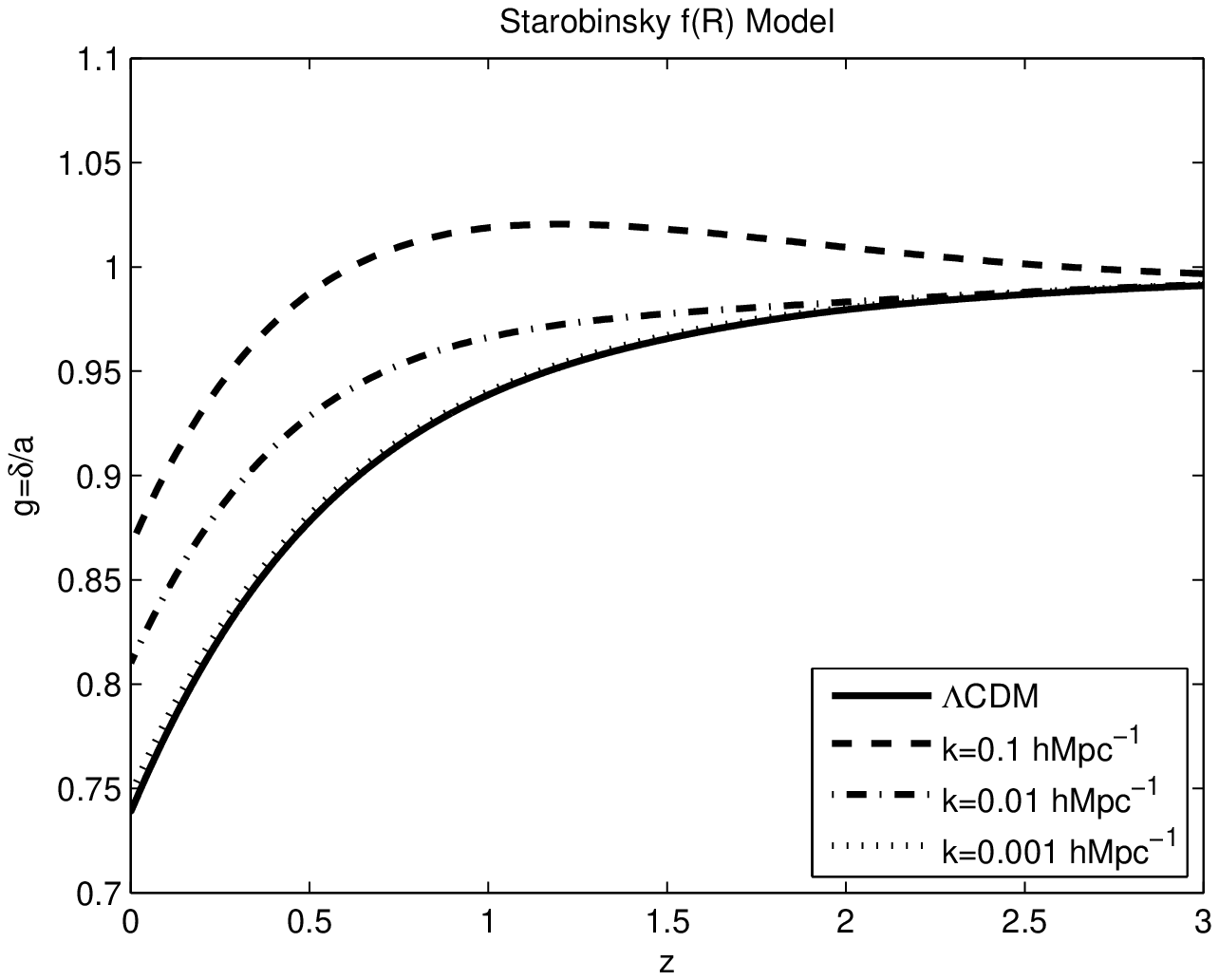}} \hspace{.1cm}
\subfigure{ \includegraphics[width=.48\textwidth]%
{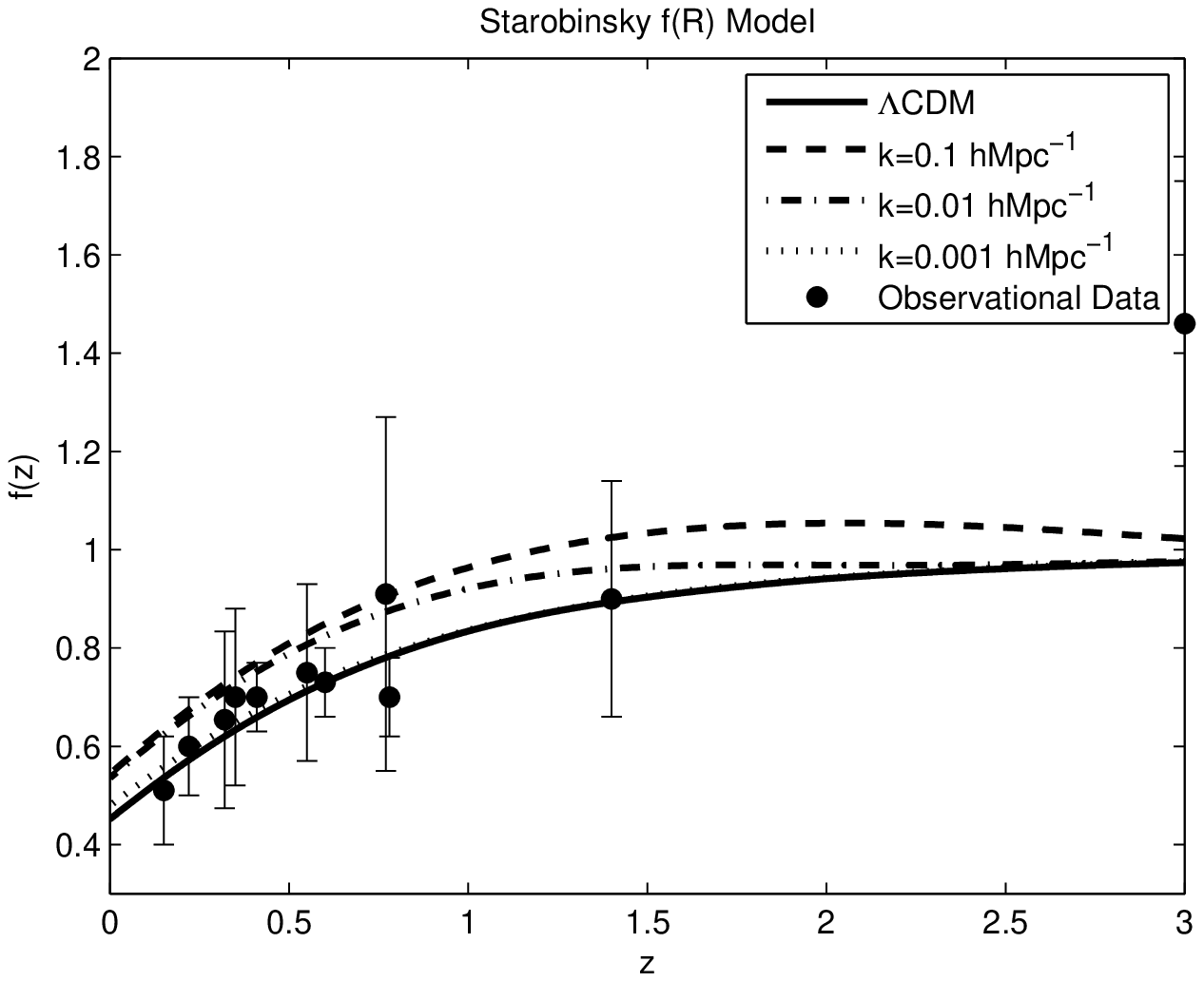}}
\end{minipage}
\caption{The variations of $RF_{\rm R}$, the screened mass function
$G_{\rm eff}/G$, the linear density contrast relative to its value
in a pure matter model
 $g=\delta/a$ and the growth factor $f(z)$, versus
redshift $z$ for the Starobinsky model.} \label{ST2}
\end{figure}

\begin{figure}[h]
\begin{minipage}[b]{1\textwidth}
\subfigure{ \includegraphics[width=.48\textwidth]%
{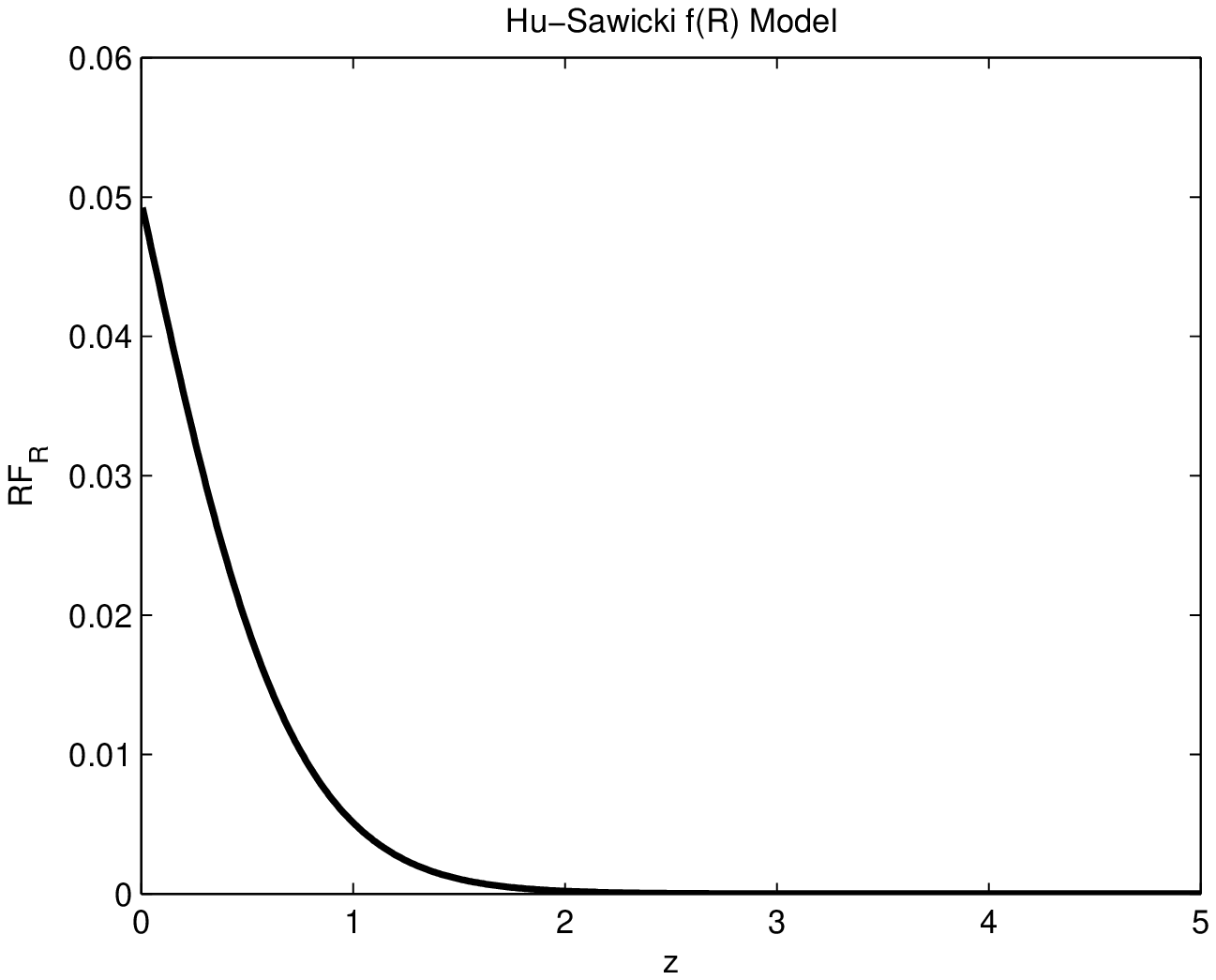}} \hspace{.1cm}
\subfigure{ \includegraphics[width=.48\textwidth]%
{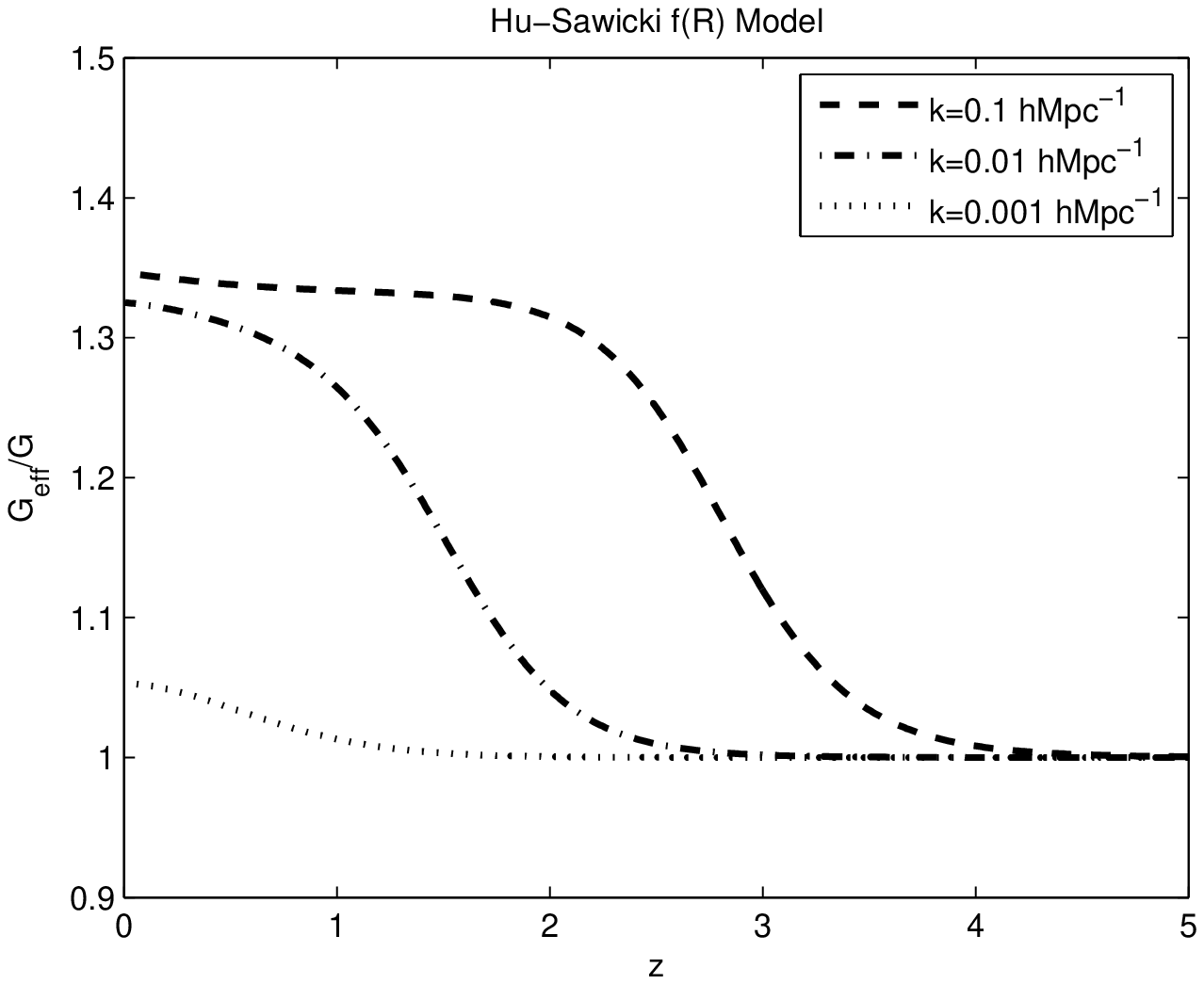}}
\end{minipage}
\begin{minipage}[b]{1\textwidth}
\subfigure{ \includegraphics[width=.48\textwidth]%
{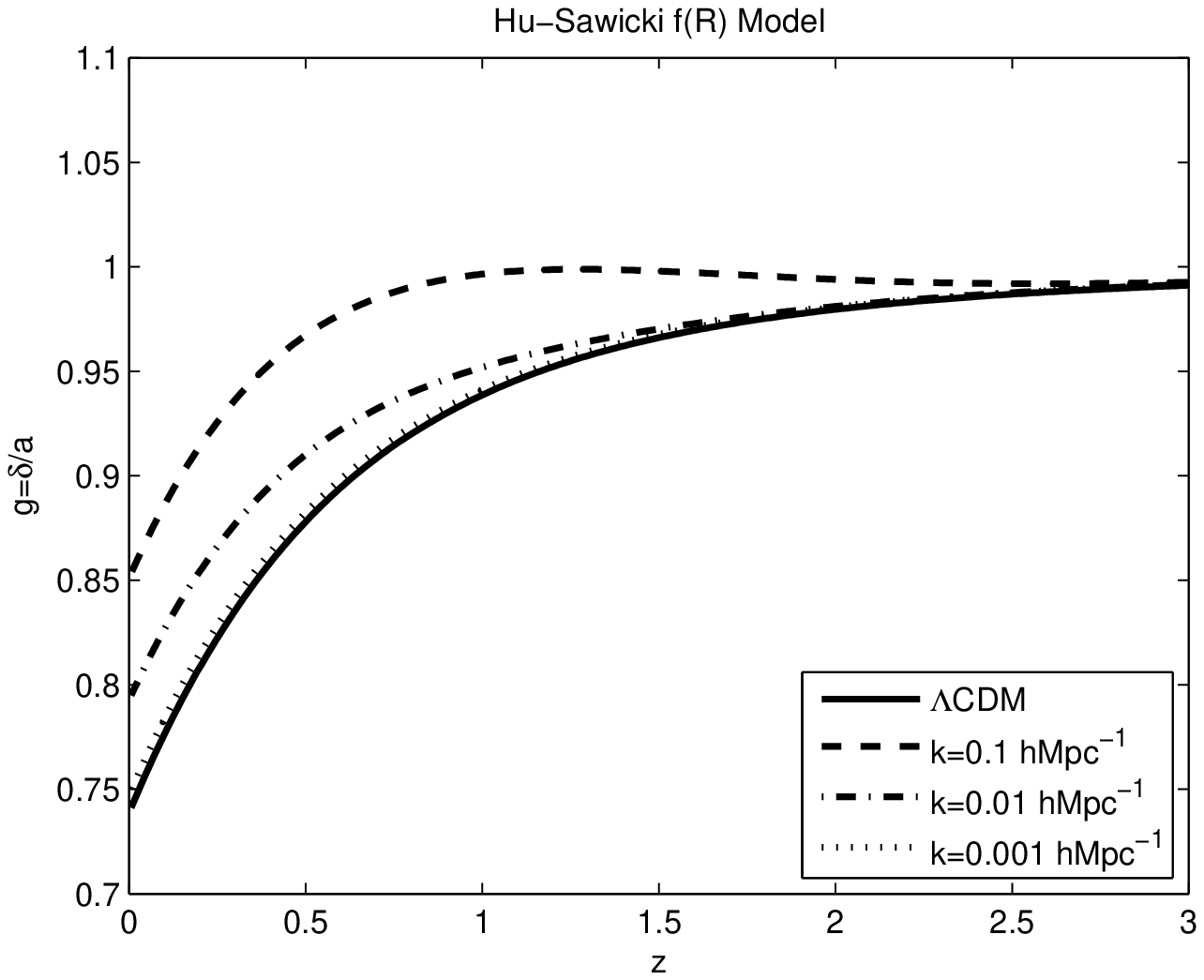}} \hspace{.1cm}
\subfigure{ \includegraphics[width=.48\textwidth]%
{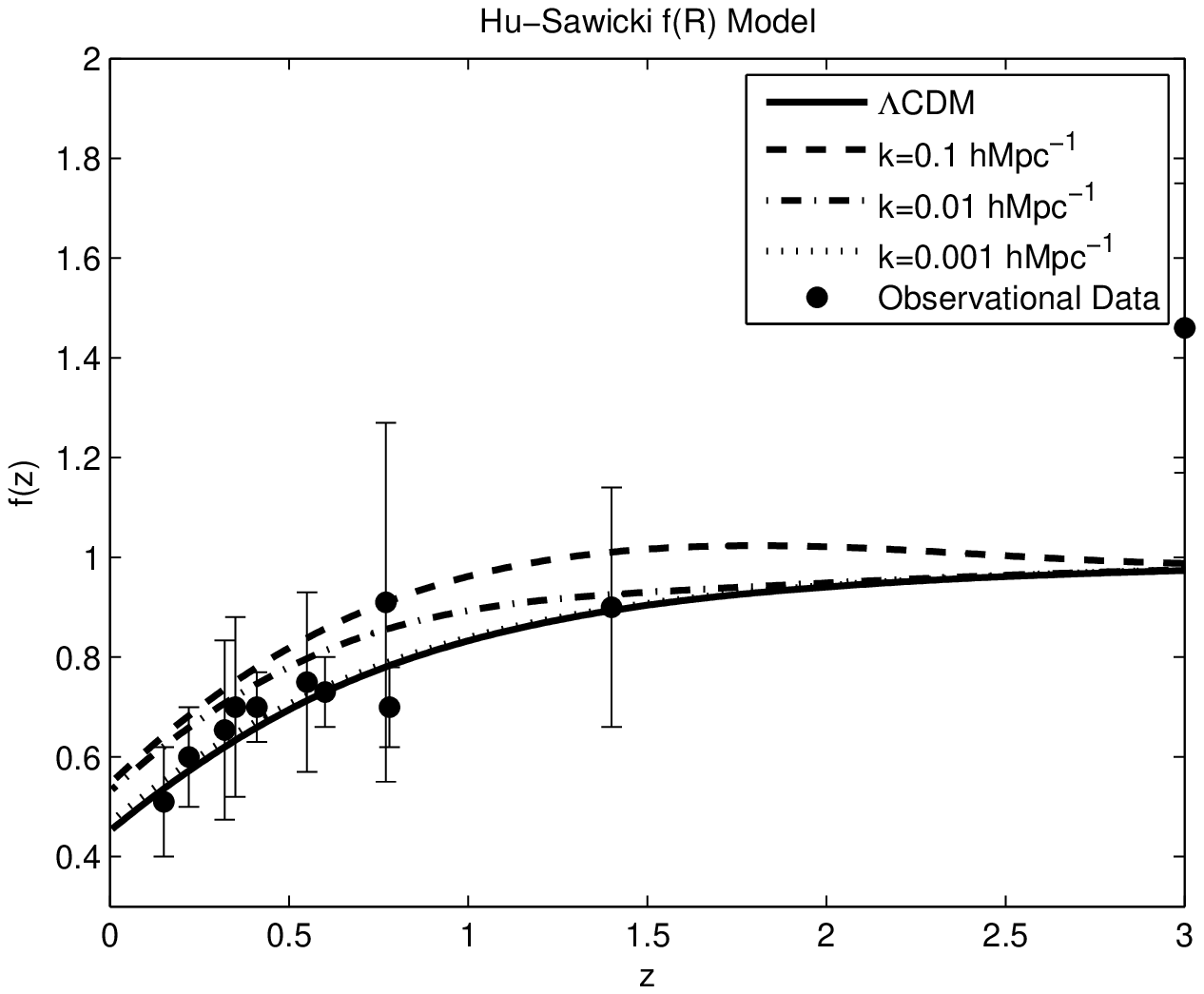}}
\end{minipage}
\caption{Same as Fig. \ref{ST2} but for the Hu-Sawicki model.}
\label{Hu2}
\end{figure}

\begin{figure}[h]
\begin{minipage}[b]{1\textwidth}
\subfigure{ \includegraphics[width=.48\textwidth]%
{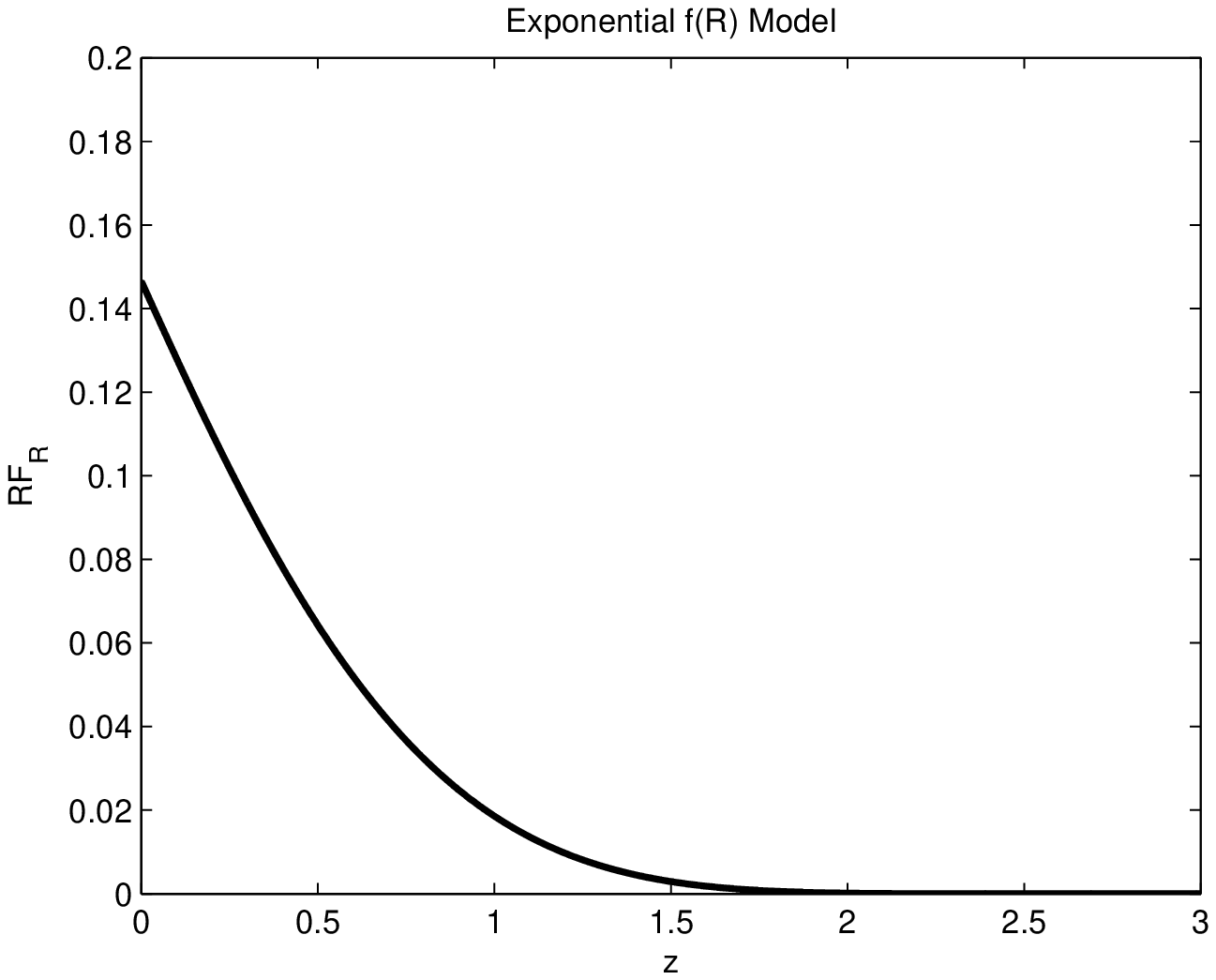}} \hspace{.1cm}
\subfigure{ \includegraphics[width=.48\textwidth]%
{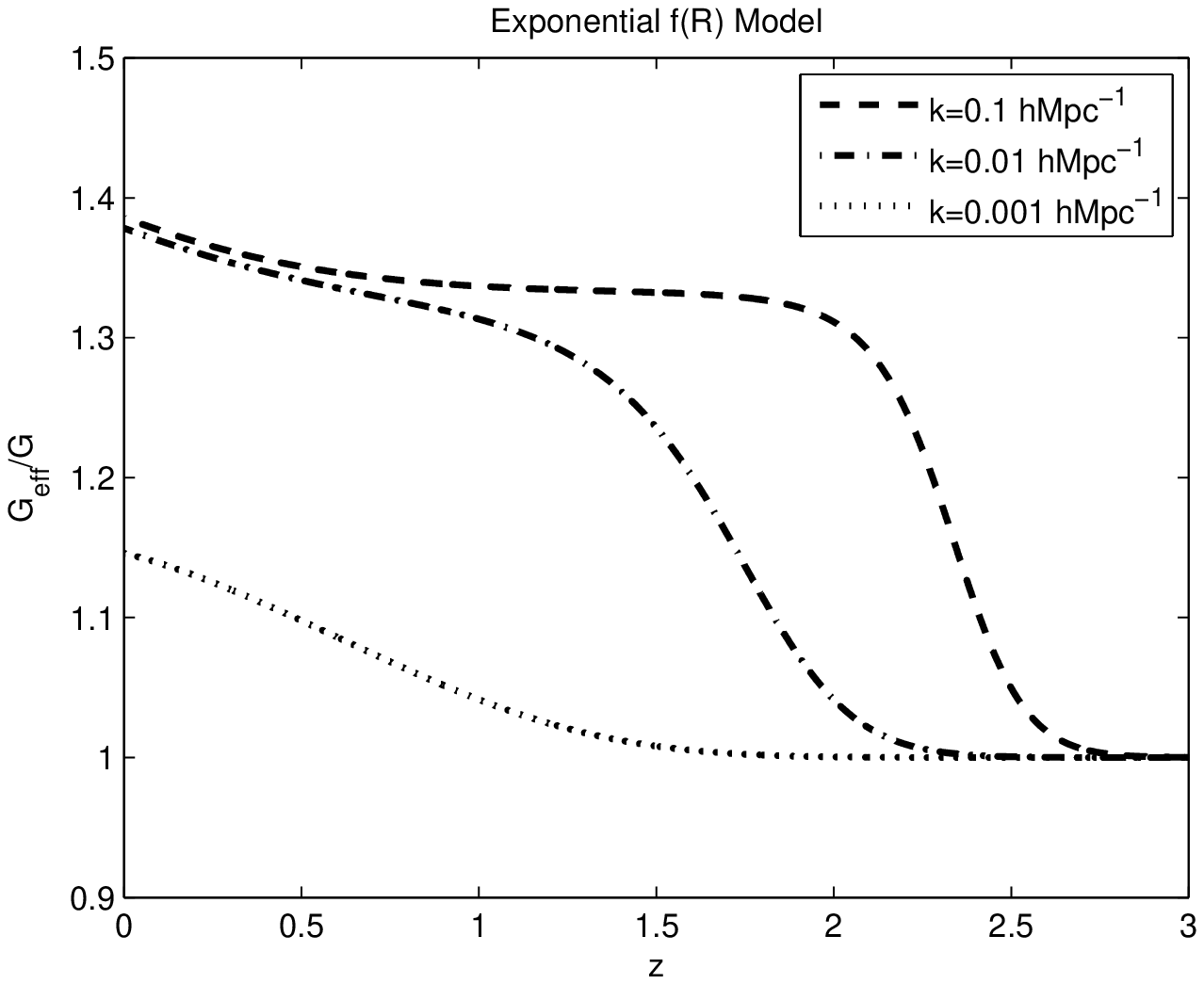}}
\end{minipage}
\begin{minipage}[b]{1\textwidth}
\subfigure{ \includegraphics[width=.48\textwidth]%
{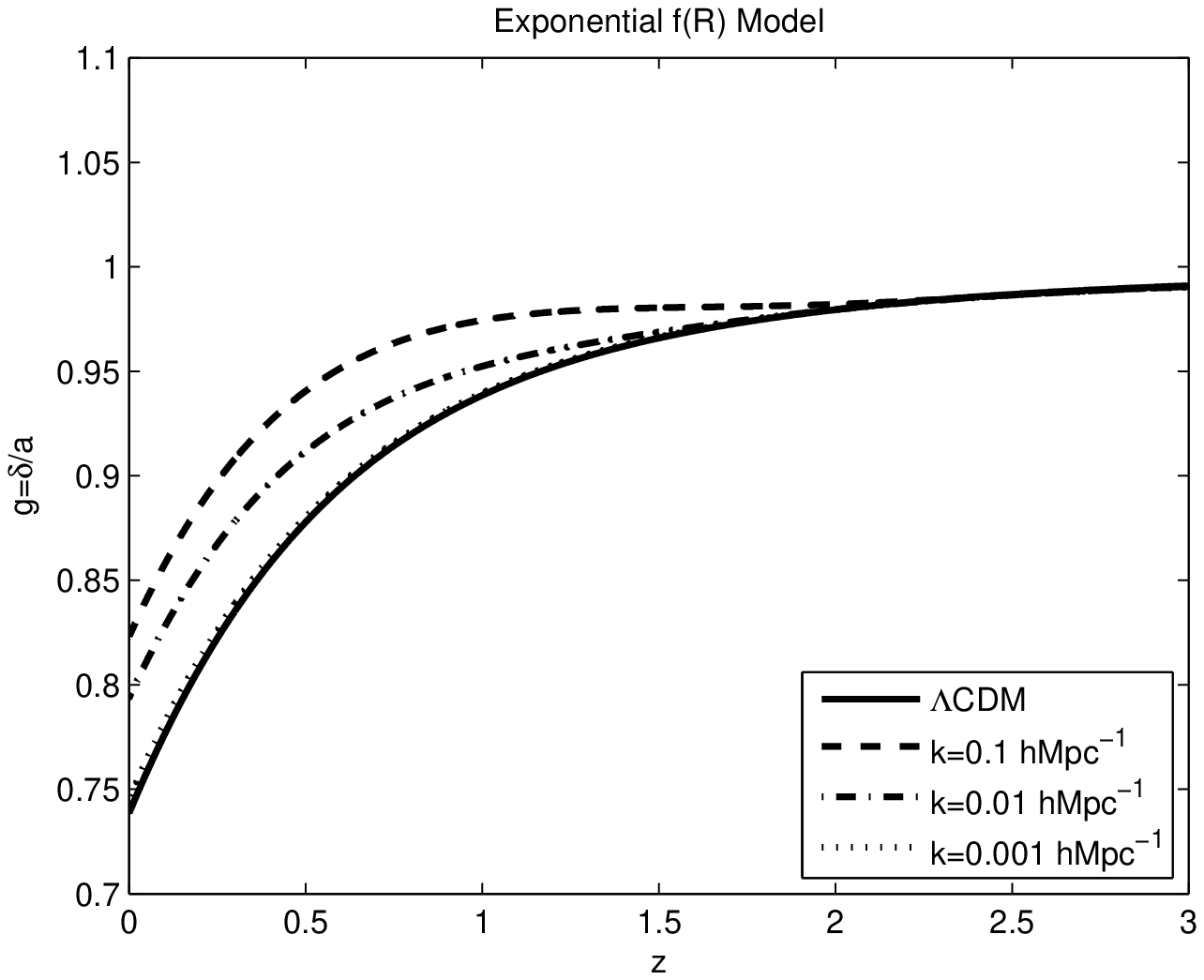}} \hspace{.1cm}
\subfigure{ \includegraphics[width=.48\textwidth]%
{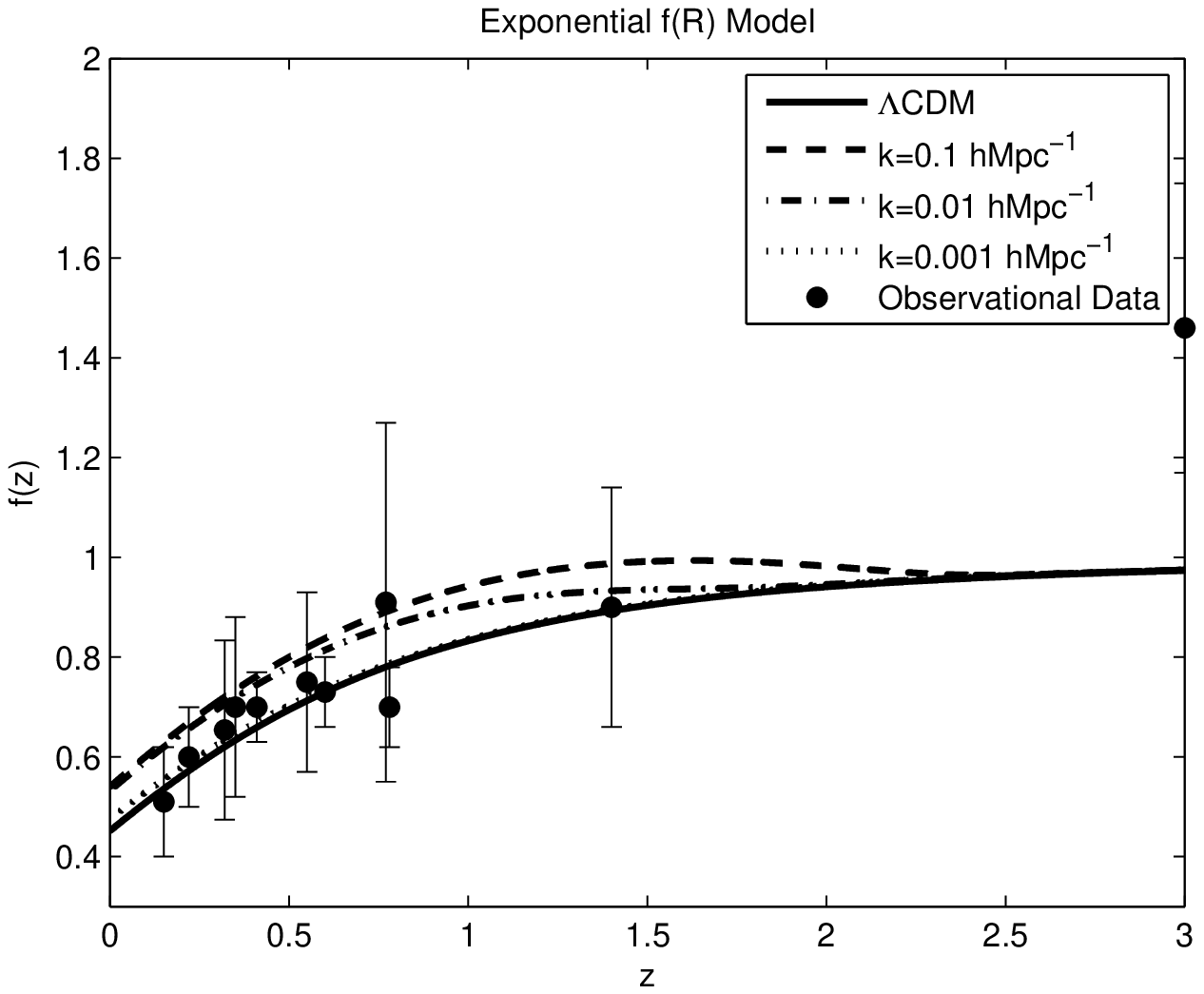}}
\end{minipage}
\caption{Same as Fig. \ref{ST2} but for the Exponential model.}
\label{Exp2}
\end{figure}

\begin{figure}[h]
\begin{minipage}[b]{1\textwidth}
\subfigure{ \includegraphics[width=.48\textwidth]%
{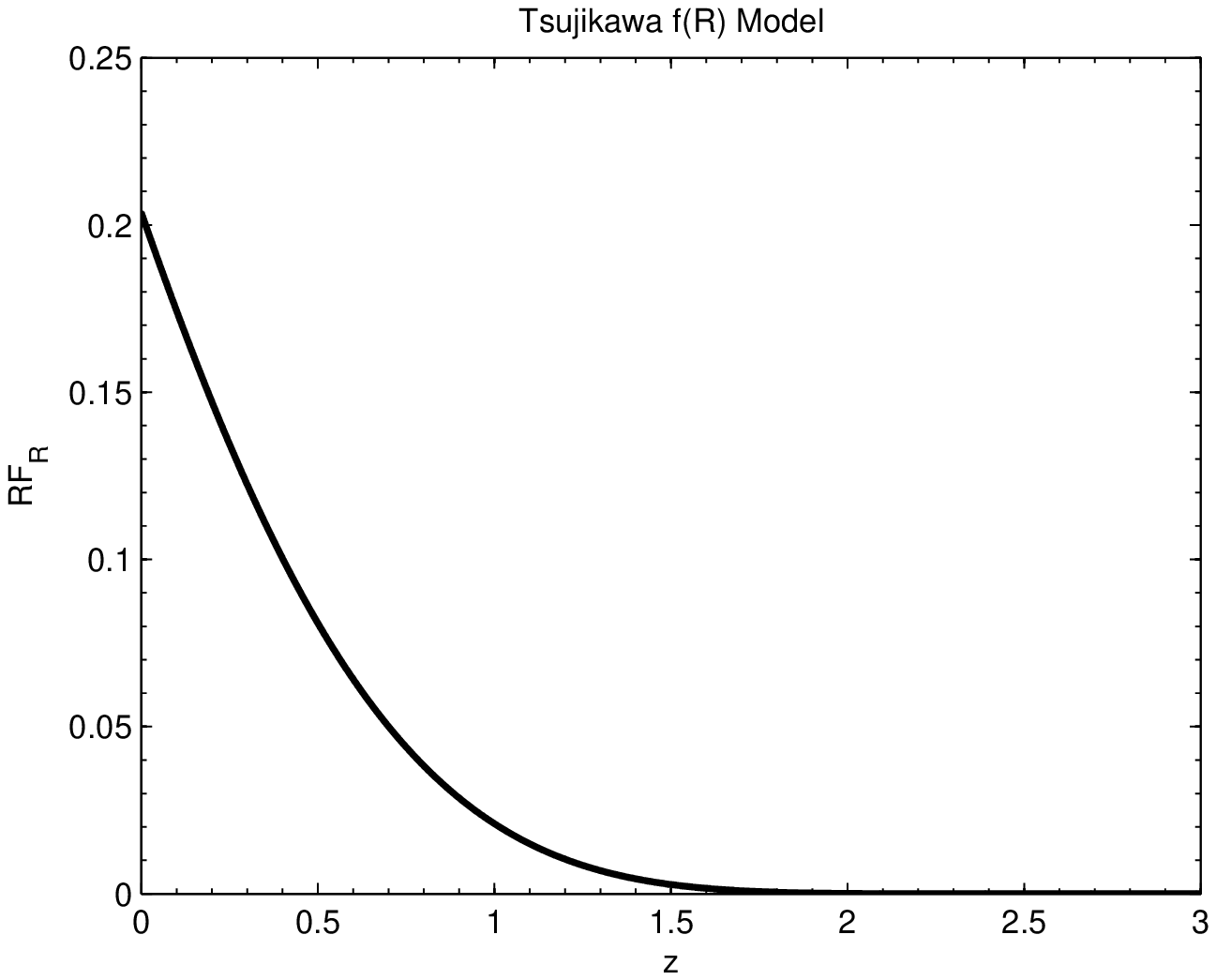}} \hspace{.1cm}
\subfigure{ \includegraphics[width=.48\textwidth]%
{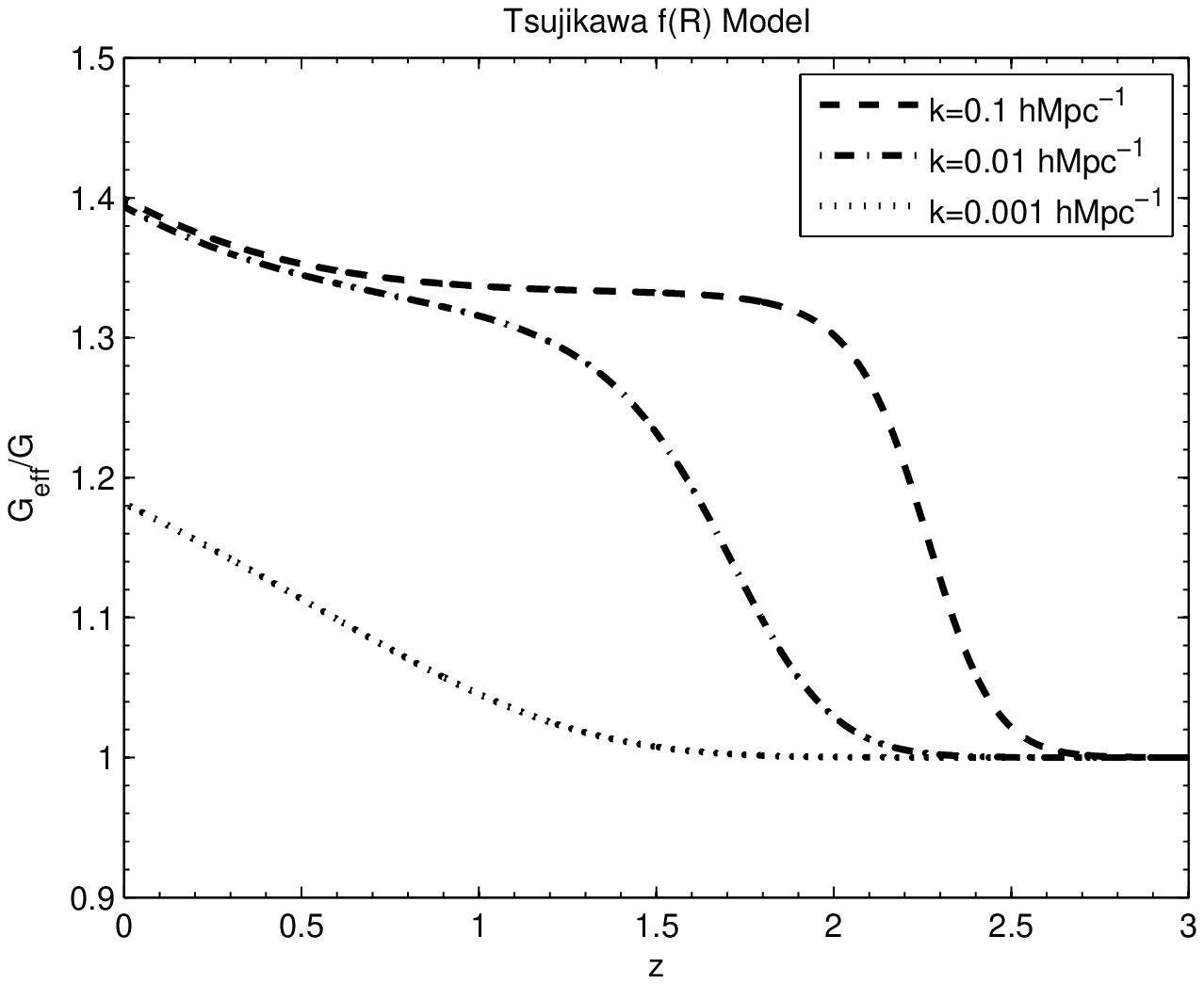}}
\end{minipage}
\begin{minipage}[b]{1\textwidth}
\subfigure{ \includegraphics[width=.48\textwidth]%
{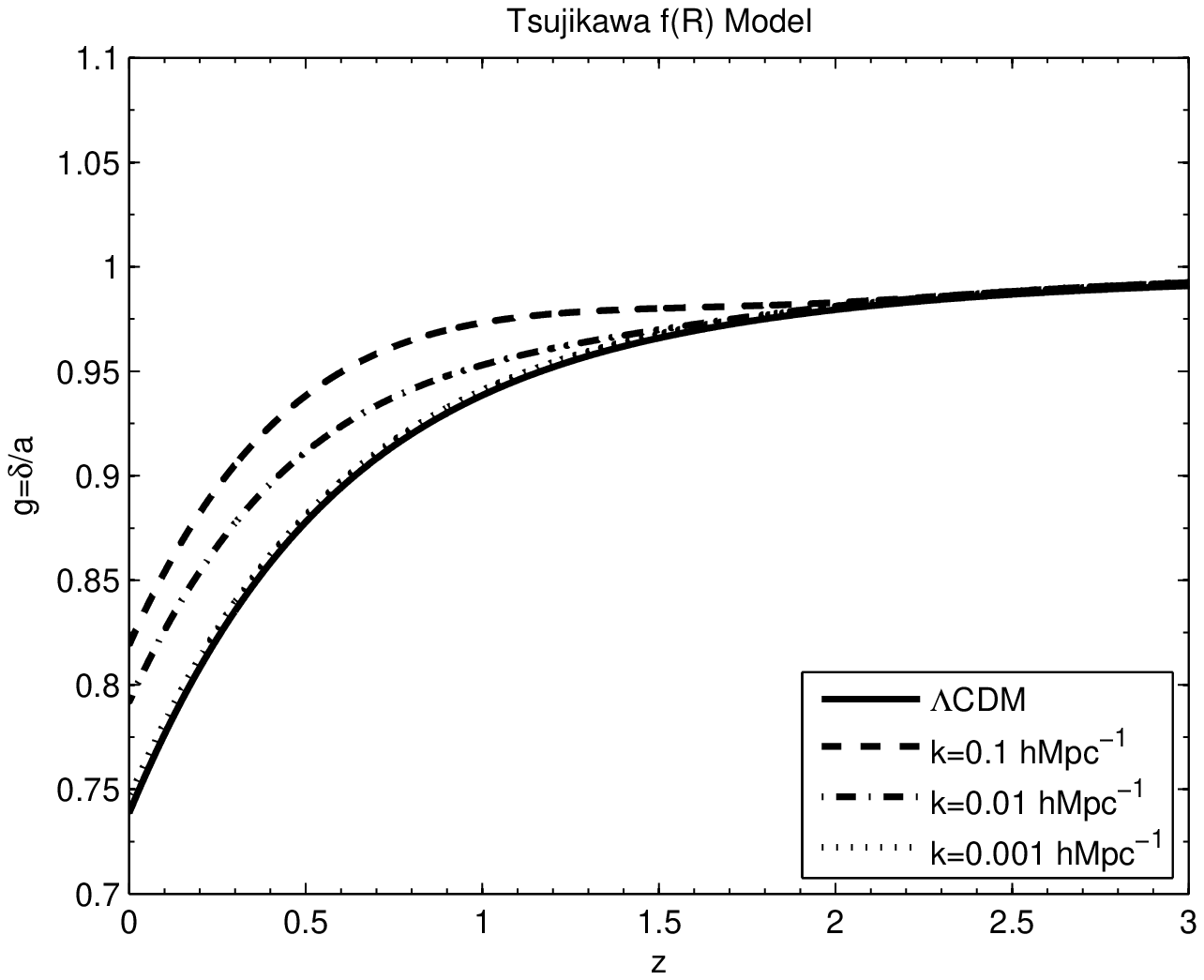}} \hspace{.1cm}
\subfigure{ \includegraphics[width=.48\textwidth]%
{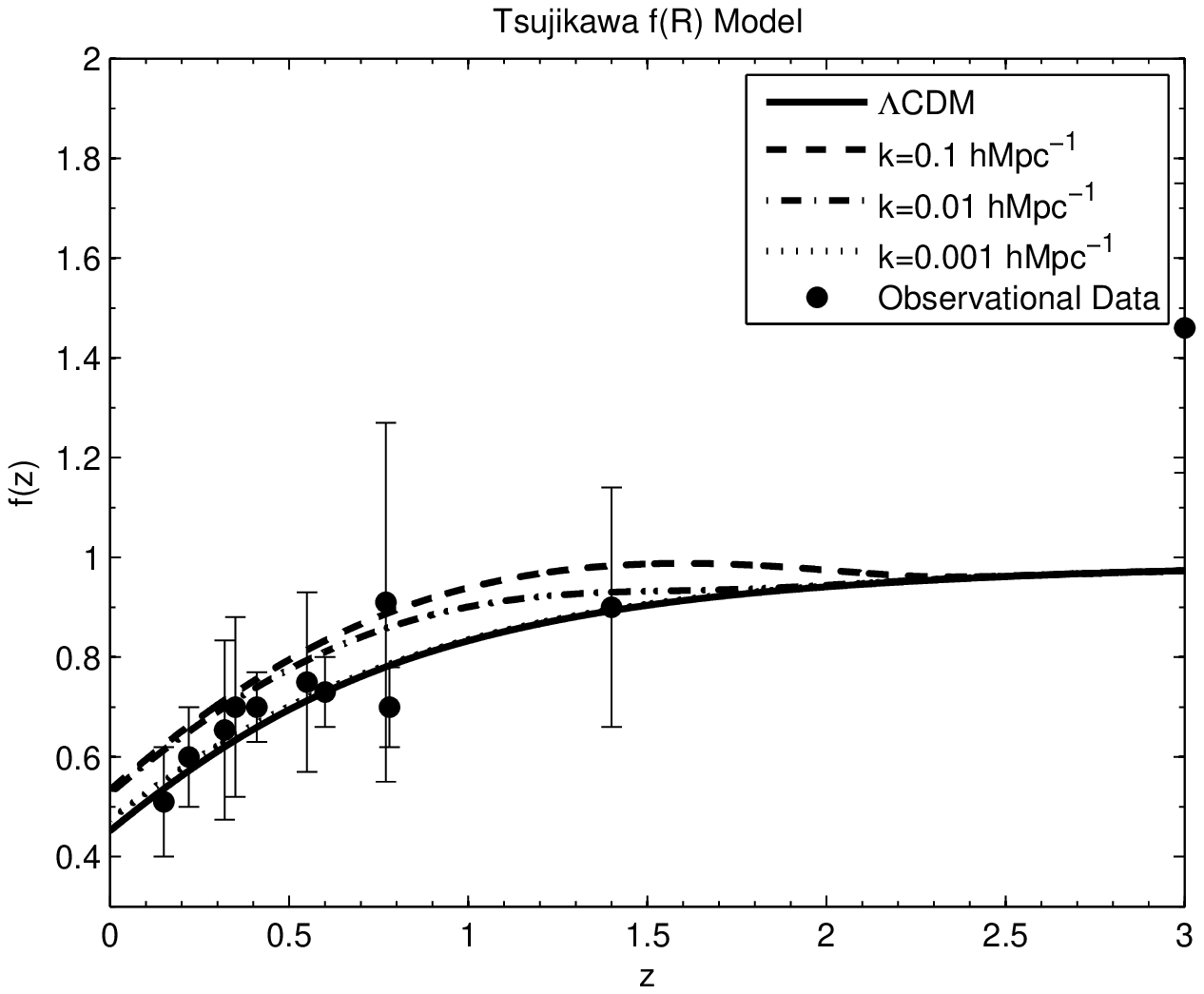}}
\end{minipage}
\caption{Same as Fig. \ref{ST2} but for the Tsujikawa model.}
\label{Tsu2}
\end{figure}

\begin{figure}[h]
\begin{minipage}[b]{1\textwidth}
\subfigure{ \includegraphics[width=.48\textwidth]%
{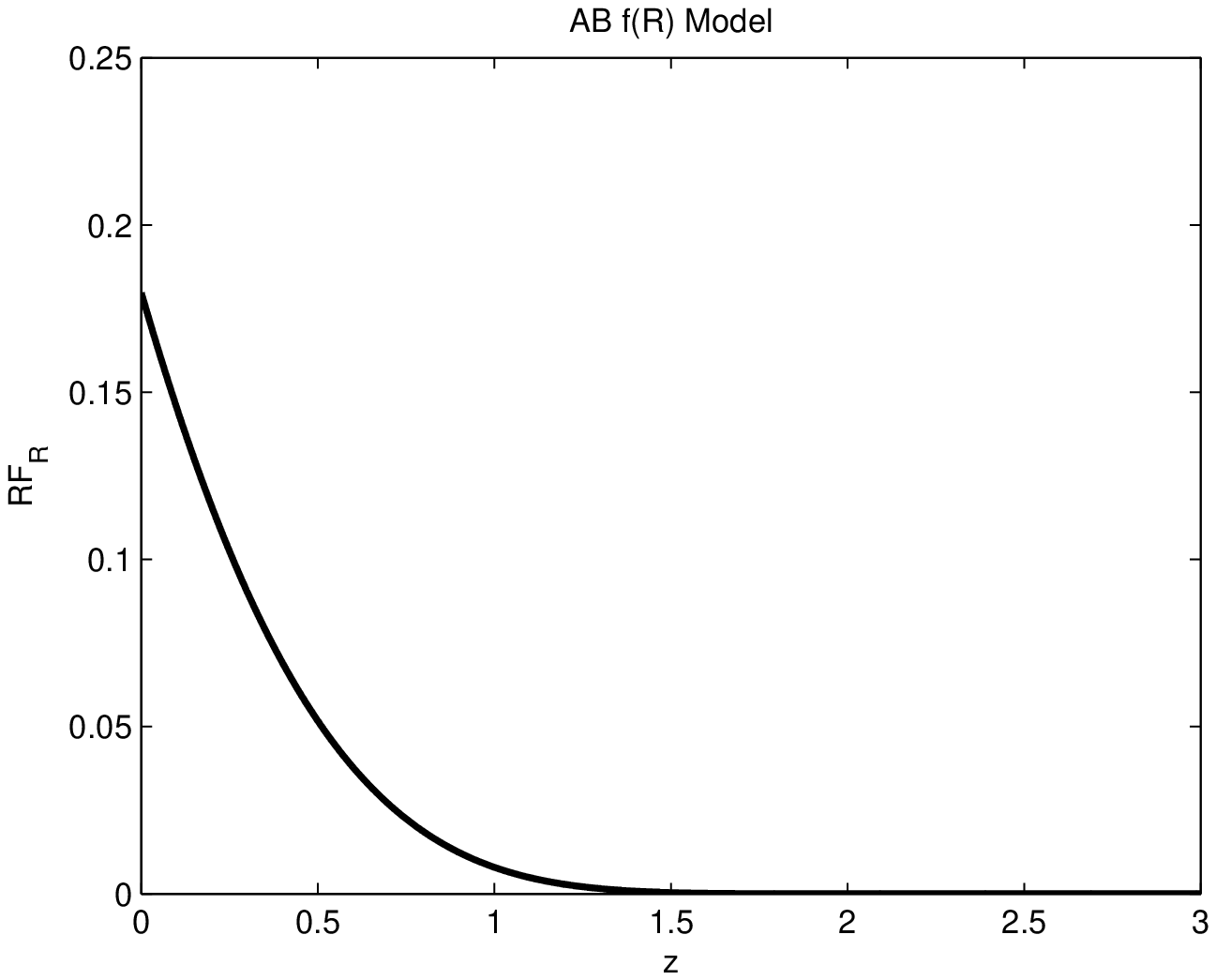}} \hspace{.1cm}
\subfigure{ \includegraphics[width=.48\textwidth]%
{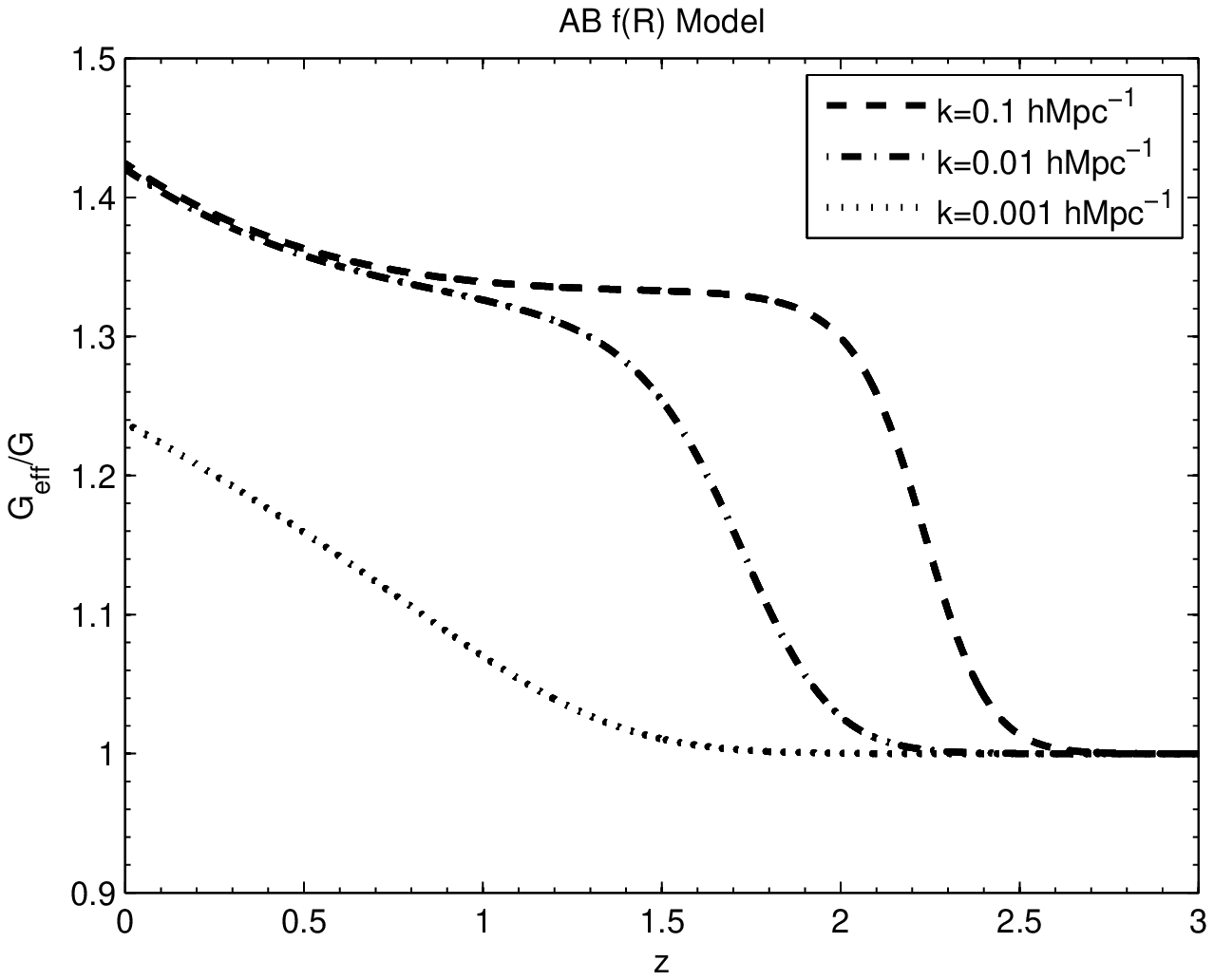}}
\end{minipage}
\begin{minipage}[b]{1\textwidth}
\subfigure{ \includegraphics[width=.48\textwidth]%
{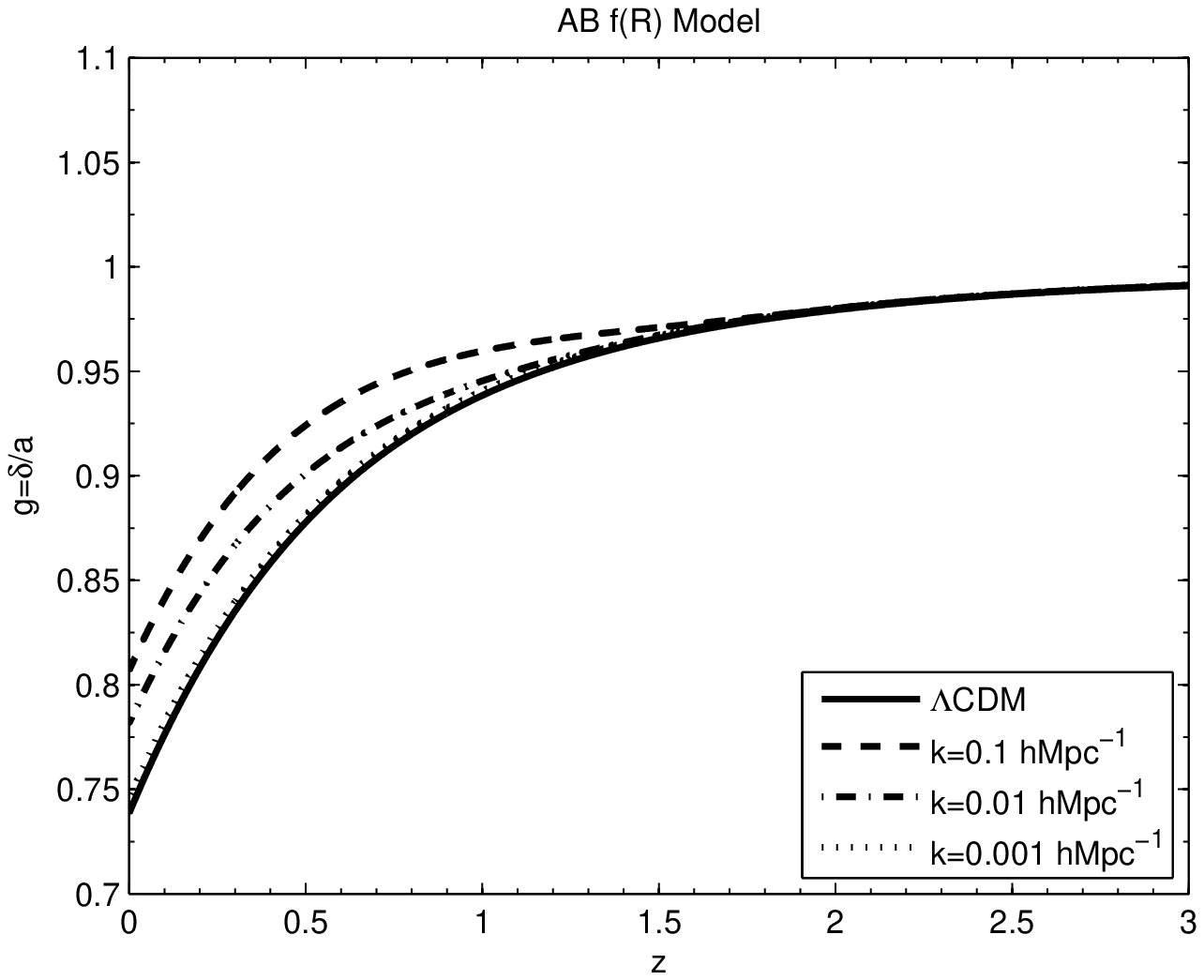}} \hspace{.1cm}
\subfigure{ \includegraphics[width=.48\textwidth]%
{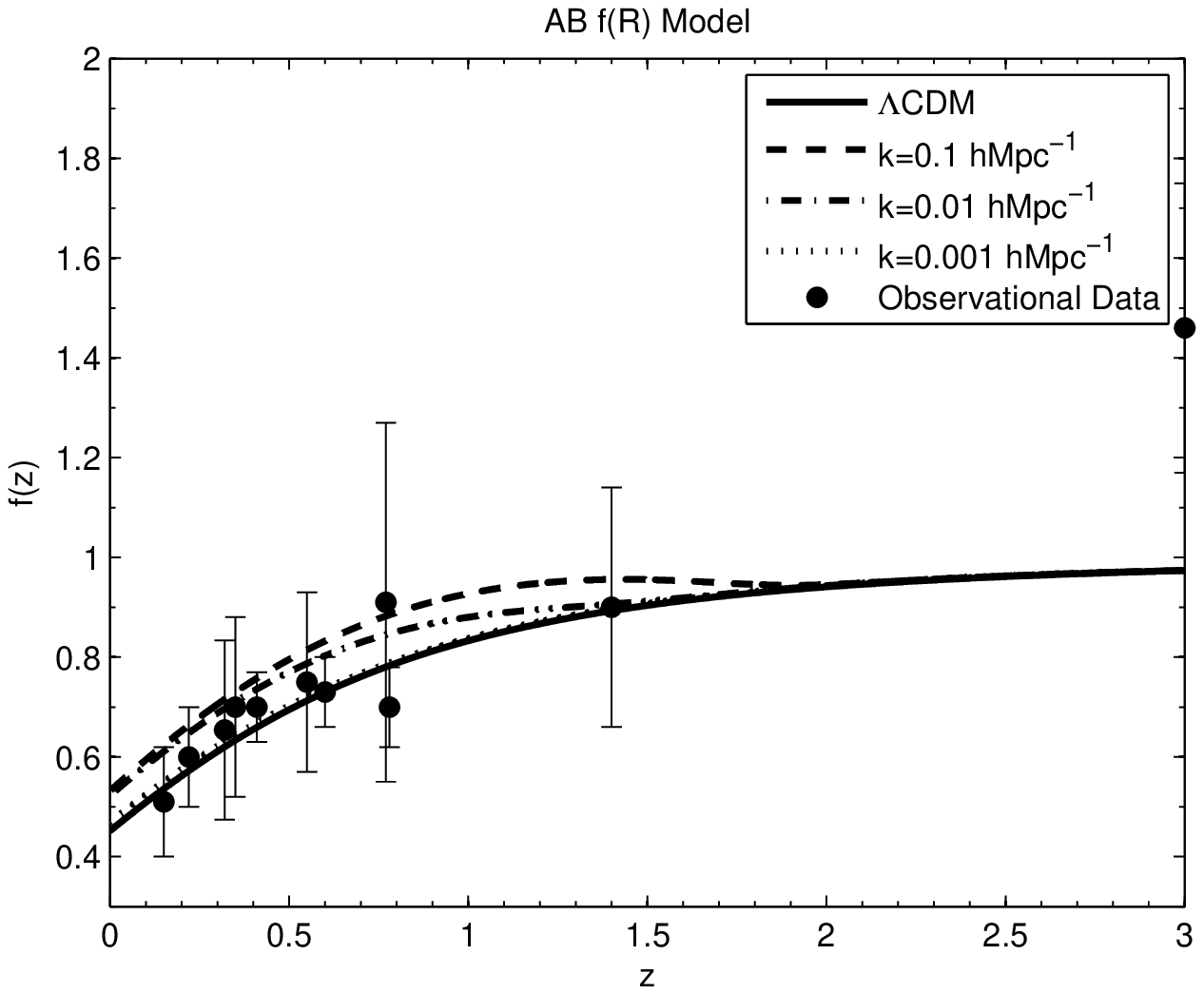}}
\end{minipage}
\caption{Same as Fig. \ref{ST2} but for the AB model.}
\label{AB2}
\end{figure}

In Figs. \ref{ST2}-\ref{AB2}, we plot the evolutions of $R F_{\rm
R}$, $G_{\rm eff}/G$, $g$ and the growth factor $f$ versus $z$ for
the selected $f(R)$ models. Figures show that: (i) $RF_{\rm R}$ goes
to zero for higher values of $z$ which means that the $f(R)$ models
at high $z$ regime behave like the $\Lambda$CDM model. (ii) The
screened mass function $G_{\rm eff}/G$ for a given wavenumber $k$ is
larger than one which makes a faster growth of the structures
compared to the GR. However, for the higher redshifts, the screened
mass function approaches to unity in which the GR structure
formation is recovered. Note that the deviation of $G_{\rm eff}/G$
from unity for small scale structures (larger $k$) is greater than
large scale structures (smaller $k$). (iii) The linear density
contrast relative to its value in a pure matter model $g=\delta/a$
starts from an early matter-dominated phase, i.e. $g\simeq1$ and
decreases during history of the universe. For a given $z$, $g$ in
the all $f(R)$ models, is greater than that in the $\Lambda$CDM
model. (iv) The evolution of the growth factor $f(z)$ for $f(R)$
models and $\Lambda$CDM model together with the 11 observational
data of the growth factor listed in Table \ref{fdata} show that for
smaller structures (larger $k$), the all $f(R)$ models deviate from
the observational data. But for larger structures (smaller $k$), the
growth factor in the all $f(R)$ models, very similar to the
$\Lambda$CDM model, fits the data very well.
\section{Conclusions}\label{con}

Here, we investigated the evolution of both matter density
fluctuations and GSL in some viable $f(R)$ models containing the
Starobinsky, Hu-Sawicki, Exponential, Tsujikawa and AB
models. For the aforementioned models, we first obtained the
evolutionary behaviors of the Hubble parameter, the Ricci scalar,
the deceleration parameter, the matter and DE density parameters,
the EoS parameters and the GSL. Then, we explored the growth of
structure formation in the selected $f(R)$ models. Our results show
the following.

(i) All of the selected $f(R)$ models can give rise to a late time
accelerated expansion phase of the universe. The deceleration
parameter for the all models shows a cosmic deceleration $q>0$ to
acceleration $q<0$ transition. The present value of the deceleration
parameter takes place in the observational range. Also at late times
($z\rightarrow -1$), it approaches a de Sitter regime (i.e.
$q\rightarrow -1$), as expected.

(ii) The effective EoS parameter $\omega_{\rm eff}$ for the all
models starts from the matter dominated era, $\omega_{\rm eff}\simeq
0$, and in the late time, $z\rightarrow -1$, it behaves like the
$\Lambda$CDM model, $\omega_{\rm eff}\rightarrow -1$.

(iii) The evolution of the EoS parameter of DE, $\omega_{\rm
D}$, shows that the crossing of the phantom divide line
$\omega_{D}=-1$ appears in the near past as well as farther future.
This is a common physical phenomena to the existing viable $f(R)$
models and thus it is one of the peculiar properties of $f(R)$
gravity models characterizing the deviation from the $\Lambda$CDM
model \cite{Bamba2}.

(iv) The GSL is respected from the early times to the present epoch.
But in the farther future, the GSL for the all models is
violated in some ranges of redshift. The physical reason why the GSL
does not hold in the farther future is that the sign of $\ddot{F}$
changes from positive to negative due to the dominance of DE over
non-relativistic matter.

(v) For the all models, the screened mass function $G_{\rm eff}/G$
is larger than 1 and in high $z$ regime goes to 1. The deviation of
$G_{\rm eff}/G$ from unity for larger $k$ (smaller structures) is
greater than the smaller $k$ (larger structures). The modification
of GR in the framework of $f(R)$-gravity, gives rise to an effective
gravitational constant, $G_{\rm eff}$, which is time and scale
dependent parameter in contrast to the Newtonian gravitational
constant.

(vi) The linear density contrast relative to its value in a pure
matter model, $g(a)=\delta_{\rm m}/a$, for the all models starts
from an early matter-dominated phase, $g(a)=1$, and decreases during
history of the universe.

(vii) The evolutionary behavior of the growth factor of linear
matter density perturbations, $f(z)$, shows that for the all models
the growth factor for smaller $k$ (larger structures) like the
$\Lambda$CDM model fit the data very well.



\clearpage

\clearpage

\clearpage

\clearpage

\clearpage

\clearpage

\clearpage

\clearpage

\clearpage

\clearpage

\clearpage

\begin{table}\small
\centering \caption{The observational data for the linear growth
rate $f_{\rm obs}(z)$.}
\begin{tabular}{lccccccccccc}\hline
$z$ & $0.15$ & $0.22$  & $0.32$ & $0.35$ & $0.41$ & $0.55$ & $0.60$
& $0.77$ & $0.78$ & $1.4$ & $3.0$
\\\hline
 $f_{\rm obs}$ &  $0.51$ &$0.60$ & $0.654$ & $0.70$ & $0.70$ & $0.75$ &
$0.73$  &$0.91$& $0.70$ & $0.90$ & $1.46$
\\\hline$1\sigma$ & $0.11$ &$0.10$ & $0.18$ & $0.18$ &
$0.07$ & $0.18$ & $0.07$ & $0.36$ & $0.08$ & $0.24$ & $0.29$
\\\hline Ref. & \cite{Hawkins} &\cite{Blake} &
\cite{Reyes} & \cite{Tegmarks} & \cite{Blake} & \cite{Ross} &
\cite{Blake} & \cite{Guzzo} & \cite{Blake} & \cite{Angela} &
\cite{Donald}\\\hline
\end{tabular}
\label{fdata}\\
\end{table}

\end{document}